\documentclass[pra,superscriptaddress,twocolumn,showpacs]{revtex4}
\usepackage{amssymb,amsmath}
\usepackage{color}
\usepackage{dsfont}
\usepackage{hyperref}
\usepackage{graphicx}

\definecolor{darkblue}{rgb}{0,0,0.5}
\definecolor{darkred}{rgb}{0.7,0,0}

\hypersetup{
  pdfborder={0,0,0},
  colorlinks=true,
  linkcolor=darkblue,
  citecolor=blue
}

\begin{document}

\title{Quantum Collapse Bell Inequalities}

\author{Karl-Peter Marzlin}
\affiliation{Department of Physics, St. Francis Xavier University,
  Antigonish, Nova Scotia, B2G 2W5, Canada}
\affiliation{Institute for Quantum Information Science,
        University of Calgary, Calgary, Alberta T2N 1N4, Canada}

\author{T.~A.~Osborn}
\affiliation{Department of Physics and Astronomy, University of Manitoba,
Winnipeg, Manitoba R3T 2N2, Canada} \bigskip

\begin{abstract}
We propose Bell inequalities for discrete or continuous quantum
systems which test the compatibility of quantum physics with an
interpretation in terms of deterministic hidden-variable theories.
The wave function collapse that occurs in a sequence of quantum
measurements enters the upper bound via the concept
of quantum conditional probabilities. The resulting hidden-variable
inequality is applicable to an arbitrary observable that is decomposable
into a weighted sum of non-commuting
projectors.
We present local and non-local examples of violation of generalized Bell
inequalities in phase space, which sense the negativity of the
Wigner function.
\end{abstract}


\maketitle

\section{Introduction}
Since its discovery in 1964 \cite{Bell:Physics1964} the Bell
inequality (BI)
has triggered an enormous interest in the differences
between classical and quantum correlations.
Bell inequalities are now commonly referred to as
relations between correlation measurements that
are fulfilled in hidden variable (HV) theories,
but are violated within the framework of quantum mechanics (QM).
The original inequality
was formulated for dichotomic variables in spin systems.
Clauser, Horne, Shimony, and Holt (CHSH)
\cite{PhysRevLett.23.880} presented a BI that
was more amenable for experimental tests
\cite{PhysRevLett.28.938,PhysRevLett.49.1804,PhysRevLett.81.5039,BellLoopholeZeilinger2013}
and is nowadays widely used.
  The latter is often formulated by introducing a `Bell operator'  $
  \hat{{\cal B}} $ and is then given by
\begin{align}\label{eq1}
&|\langle    \hat{{\cal B}} \rangle_\text{QM}  | \leq 2 \\
  &\hat{{\cal B}} = \hat{A}\otimes \hat{B}+ \hat{A}\otimes\hat{B}'+ \hat{A}' \otimes\hat{B} -
  \hat{A}' \otimes\hat{B}' , \quad
\label{eq:CHSHBellOp}\end{align}
where $\langle    \hat{{\cal B}} \rangle_\text{QM} $ is the quantum
expectation value of operator $ \hat{{\cal B}} $.
The operators $\hat{A},\hat{A}'$ and $\hat{B},\hat{B}'$ are
dichotomous
 (i.e, they have only two eigenvalues)
and act on different quantum systems A and B, respectively.

The original BI was inspired by the Einstein-Podolsky-Rosen paradox
\cite{PhysRev.47.777,RevModPhys.81.1727} for
 infinite-dimensional quantum systems, but
BIs for such systems were developed much later. The first
proposals also used dichotomic observables
\cite{JModOpt42-939,Gour2004415,Praxmeyer:EPJD2005}, but recently
a new approach has been
developed by Cavalcanti, Foster, Reid, and Drummond (CFRD)
\cite{PhysRevLett.99.210405,arXiv:1005.2208}. The CFRD inequality
is based on an argument  that involves HV commutativity and
can be formulated for arbitrary quantum systems.

In this paper we
propose an alternative approach to BIs for infinite
  systems, which can be applied to an arbitrary observable
and provides explicit links between HV theories and the corresponding
quantum system. Our derivation is
based on the decomposition of a general Bell operator
 $  \hat{{\cal B}} $ into a superposition of projectors.
The BI makes essential use of  wave function collapse expressed via
quantum conditional probabilities.

In Sec.~\ref{sec:BImain} we will present the main result and discuss
its features.
In Sec.~\ref{sec:chsh} we demonstrate that the
proposed inequality is consistent with the CHSH inequality
for a Bell operator of the form (\ref{eq:CHSHBellOp}).
We derive a generic form of the generalized BI in phase space in
Sec.~\ref{sec:quant} and subsequently give examples of its violation
for single-particle (Sec.~\ref{sec:loc})  and bi-partite quantum systems (Sec.~\ref{sec:nonloc}).
Several appendices contain the details of our derivations.

\section{Generalized Bell inequalities}\label{sec:BImain}

We consider a general Bell operator $\hat{{\cal B}}$ acting on a
 generic Hilbert
space $\cal{H}$ of ${\dim(\cal{H})} \ge 3 $ that allows both continuous and discrete (spin) degrees of freedom.
The only feature required  of this revised Bell operator is that it can be
decomposed into a set of projectors $\hat{P}(u) $ as
\begin{align}
  \hat{{\cal B}} &= \int \hspace{-4mm} {\textstyle \sum} du\,  w(u) \hat{P}(u) .
\label{eq:BellOpDecomp} \end{align}
In this expansion, $u$ may represent a set of several
variables and the symbol $ \int \hspace{-3.3mm} {\scriptstyle  \sum }du$
denotes a sum, an integral, or a combination of both, over the
variables represented by $u$. The weight factors $w(u)$ are real. The projectors $\hat{P}(u)$
correspond to the observables that are measured in an experiment.
In this way the experimental configuration selects the family of non-commuting projectors that appear in (\ref{eq:BellOpDecomp}).
To incorporate an
experimentally accessible form of locality in a bi-partite system, the projectors   need to be of
 tensor product form.
They then play the same role as (projectors onto eigenstates of)
the observables $\hat{A},\hat{A}', \hat{B},\hat{B}'$
of Eq.~(\ref{eq:CHSHBellOp}). In Sec.~\ref{sec:chsh} we will
make this connection explicit.

In formulating generalized BIs we utilize the state after
a measurement of observable $\hat{P}(u)$ has been performed.
Let $\rho$ denote the initial density matrix
of a quantum system.
After a measurement of projector $\hat{P}(u) $ has been
performed, the state will collapse to
\begin{align}
  \rho_{u}&\equiv
 \frac{ 1}{\text{Tr}(\rho \hat{P}(u) )}   \hat{P}(u)\, \rho \hat{P}(u).
\end{align}
Our main result can then be stated as follows.
\\[3mm]
{\bf Theorem 1}: Let $\langle \hat{{\cal B}} \rangle_{\text{QM}} $ denote the mean
value of the   generalized Bell operator in quantum theory.
For $\langle \hat{{\cal B}} \rangle_{\text{QM}} $ to be consistent with a
deterministic
HV description, it must obey the inequality
\begin{align}
   |\langle \hat{{\cal B}} \rangle_{\text{QM}} |^2 &\leq
   \langle \hat{{\cal B}}^2  \rangle_\text{HV}
\label{eq:genBI}
\\
  \langle \hat{{\cal B}}^2 \rangle_\text{HV} &=
   \int \hspace{-4mm} {\textstyle \sum} du\,  w(u)
  \text{Tr}\left (\rho_{u}\hat{{\cal B}} \right )
   \text{Tr}\left (\rho \hat{P}(u) \right ) .
\label{eq:HVupperBound}\end{align}
The proof and the assumptions made in a deterministic
HV framework are described in
Appendix~\ref{app:proofBI}.
The right-hand side of Eq.~(\ref{eq:HVupperBound}) is quadratic in the
weight factors $w(u)$. The HV upper bound has an unusual format in that its
value is determined solely by quantum quantities.  This is possible because of the equality between HV and
quantum conditional probabilities, {\it cf.\,} Eq.~(\ref{eq:condProbRule}).

Equation~(\ref{eq:genBI}) has a simple physical interpretation: if in an
experiment $ |\langle \hat{{\cal B}} \rangle_{\text{QM}} |$ is derived from
measurements of the observables $ \hat{P}(u)$, then
the maximum value of
$ |\langle \hat{{\cal B}} \rangle_{QM}|$
that is consistent with an HV
model is given by the sum over mean values of $\hat{{\cal B}}$
in the states that are obtained after  $ \hat{P}(u)$ has been
measured. The weight factor for each measurement is given by
$w(u)$ times the probability
$\text{Tr} (\rho \hat{P}(u)  )$ to find the system in state
$\hat{P}(u)$. An alternative interpretation can be given
by expanding the operator $\hat{{\cal B}} $ in
Eq.~(\ref{eq:genBI}), which yields

\begin{align}
   |\langle \hat{{\cal B}} \rangle_{\text{QM}} |^2 \leq
   \int \hspace{-4mm} {\textstyle \sum} du\, dv\,
   w(u)\, w(v)
  \text{Tr}\left ( \hat{P}(u) \rho \hat{P}(u)   \hat{P}(v) \right ).
\label{eq:genBI2}\end{align}
The HV upper bound is then a weighted double sum of correlations
between the observables $\hat{P}(u) $ and  $\hat{P}(v)$ that
are related to the probability to measure  $\hat{P}(v)$ provided
 $\hat{P}(u)$ has been measured first.

It is instructive to compare the HV upper bound to the upper bound in
quantum physics. In Appendix~\ref{app:proofBI} we show that
\begin{align}
  \langle \hat{{\cal B}}^2\rangle_\text{QM} &=
  \langle \hat{{\cal B}}^2  \rangle_\text{HV} +
   \int \hspace{-4mm} {\textstyle \sum} du\, dv\,
   w(u)\, w(v)
\nonumber \\ &\hspace{2cm} \times
  \text{Tr}\left (\rho  \hat{P}(u)[ \hat{P}(u) , \hat{P}(v)] \right ) .
\label{eq:boundsDifference}\end{align}
Depending on the choice of observables and the quantum state,
the difference $ \langle \hat{{\cal B}}^2
\rangle_\text{QM}
-\langle \hat{{\cal B}}^2  \rangle_\text{HV} $
between the two bounds may be positive or negative.
If it is negative, a BI violation will not  occur.
Furthermore, Eq.~(\ref{eq:boundsDifference}) demonstrates
that for any difference to occur, non-commuting observables
are necessary. This is in agreement with the general results
found by Malley and Fine \cite{PhysRevA.69.022118,Malley200551}.

Not all choices of $\hat{{\cal B}} $ and all decompositions
of it will lead to a BI violation. As result (\ref{eq:boundsDifference}) shows, at least
some of the projectors must not commute.
So spectral expansions of Hermitian operators do not lead to a BI
violation.

Another example for which no BI violation is possible is the choice
$\hat{{\cal B}} =\hat{I} $. From Eq.~(\ref{eq:HVupperBound})
it then follows that
\begin{align}
   \langle \hat{{\cal B}}^2 \rangle_\text{HV} &=
     \int \hspace{-4mm} {\textstyle \sum} du\,  w(u)
   \text{Tr}\left (\rho \hat{P}(u) \right )
\\ &=
   \text{Tr}\left ( \rho  \int \hspace{-4mm} {\textstyle \sum} du\,  w(u)
   \hat{P}(u) \right )
\; = 1,
\label{eq:unitDecomp}\end{align}
so that  $ |\langle \hat{{\cal B}} \rangle_{\text{QM}} |^2=
 \langle \hat{{\cal B}}^2  \rangle_\text{HV} =1$
for any choice of decomposition.

The decomposition (\ref{eq:BellOpDecomp}) does not restrict the choice of $\hat{{\cal B}}$,
but finding a BI violation amounts to finding a suitable
combination of $\rho, \hat{{\cal B}}$ and $\{\hat P(u)\}$. This will be the topic of the
following sections.

\section{  CHSH compatibility }\label{sec:chsh}
We first demonstrate that the generalized
BI (\ref{eq:genBI}) is consistent with the CHSH inequality for
dichotomic observables. To do so we decompose each of the
operators $\hat{A},\hat{A}', \hat{B},\hat{B}'$ appearing in
Eq.~(\ref{eq:CHSHBellOp}) in the form $\hat{A}=\hat{P}^{A}_{1}
-\hat{P}^{A}_{-1}$, where $\hat{P}^{A}_{i}$ are projectors onto
eigenstates of operator $\hat{A}$ with eigenvalue $i=\pm 1$.
The Bell operator of Eq.~(\ref{eq:CHSHBellOp}) can then be written
in the form of Eq.~(\ref{eq:BellOpDecomp}),
with $ \int \hspace{-3.3mm} {\scriptstyle  \sum }du$ representing a sum over
16 terms. Explicitly, the set of all 16 projectors $\hat{P}(u),
u=1,\cdots , 16$ is given by
\begin{align}
  \begin{array}{cccc}
 \hat{P}_1^A\otimes \hat{P}_1^B, &
  \hat{P}_{-1}^A\otimes \hat{P}_{-1}^B, &
 \hat{P}_1^A\otimes  \hat{P}_1^{B'},&
  \hat{P}_{-1}^A\otimes \hat{P}_{-1}^{B'},
\\[2mm]
  \hat{P}_1^{A'}\otimes \hat{P}_1^B,&
  \hat{P}_{-1}^{A'}\otimes \hat{P}_{-1}^B,&
  \hat{P}_1^{A'}\otimes \hat{P}_{-1}^{B'},&
  \hat{P}_{-1}^{A'}\otimes\hat{P}_1^{B'},
 \\[2mm]
\hat{P}_1^A\otimes \hat{P}_{-1}^B,&
  \hat{P}_{-1}^A\otimes \hat{P}_1^B,&
 \hat{P}_1^A\otimes\hat{P}_{-1}^{B'},&
  \hat{P}_{-1}^A\otimes \hat{P}_1^{B'},
 \\[2mm]
  \hat{P}_1^{A'}\otimes \hat{P}_{-1}^B,&
  \hat{P}_{-1}^{A'}\otimes \hat{P}_1^B,&
  \hat{P}_1^{A'}\otimes \hat{P}_1^{B'},&
  \hat{P}_{-1}^{A'}\otimes \hat{P}_{-1}^{B'},
           \end{array}
\end{align}
where the weight factor $w(u)$ is equal to $+1 (-1)$ for the
first (last) eight projectors, respectively.
It is then a straightforward but tedious task to verify that
\begin{equation}\label{N}
    \sum_{u,v} \omega(u)\omega(v) \hat{P}(u) \hat{P}(v) \hat{P}(u) = 4\,  \hat I_A \otimes \hat I_B\,.
\end{equation}
As a consequence the
upper bound (\ref{eq:HVupperBound}) takes the value
$\langle \hat{{\cal B}}^2 \rangle_\text{HV}  = 4$.
Hence, the CHSH inequality
may be considered as a special case of Eq.~(\ref{eq:genBI}).

We remark that a key feature of the CHSH inequality  needed
to implement
{\em local} HV models is that all projectors $\hat{P}(u)$ have a
product structure of the form
$\hat{P}(u) = \hat{P}_A\otimes \hat{P}_B$. Because it is the
observables $\hat{P}(u)$ that should be measured in an experiment,
the product structure ensures that for a bi-partite quantum system
with two subsystems A, B one can test violation of BI by performing
local measurements on each subsystem. For space-like separated
systems, the principle of Einstein causality then ensures that
changes in the measurement settings of system A cannot have an
influence on measurements on system B and vice versa \cite{PhysRevLett.49.1804}.
In Sec.~\ref{sec:nonloc} we will show that a product structure can
also be achieved for generalized BIs.

\section{Bell inequalities in phase space }\label{sec:quant}
One of the motivations behind this work is to study Bell inequalities
in phase space, which is a natural tool to compare classical
and quantum dynamics.
Phase space methods have been used to study specific
implementations of the CHSH inequality for dichotomic operators
\cite{PhysRevLett.82.2009,PhysRevA.58.4345,revzen:022103,de2009entanglement}. 
Because of the restriction to dichotomic operators, the resulting
inequalities are conceptually similar to BI on discrete Hilbert spaces.

The best known example of a quantum phase space
description is the Wigner function $W(x)$ \cite{PhysRev.40.749}, which
depends on the phase space variable $x\equiv (q,p)$.
It can be considered as a part of the Weyl symbol calculus \cite{Foll89}.
For a single particle in one
spatial dimension, described through the usual Hilbert
space ${\cal H} =L^2(\mathds{R},\mathds{C},dq)$ of
complex, square-integrable
wave functions of a single real variable $q$,
the Wigner function can be expressed as
$ W(x) = (2\pi \hbar)^{-1}  \text{Smb}[\rho](x)$.
Here $\text{Smb}[\hat{A}]$
denotes the Weyl symbol of an arbitrary operator $\hat A $ and is defined by
\begin{align}
    \text{Smb}[\hat{A}](x) &=
  2\pi \hbar\, \text{Tr}(\hat{A} \hat{\Delta}(x))
\label{eq:wignerfunction}
\\
   \hat{\Delta}(x) &=
  \int \frac{ dq'}{2\pi\hbar}   e^{-\frac{ i}{\hbar} p q'}
   \Big| q -\frac{1}{2} q' \Big\rangle \Big\langle q +\frac{ 1}{2} q'\Big|\,.
\label{Q1}\end{align}
 For Hermitian operators
such as the density matrix the Weyl symbol is real.
The quantizer $ \hat{\Delta}(x)$ is a unitary operator \cite{Roy77,Gro76,KO3}
that enables one to transfer
the Hilbert space representation of QM to an equivalent phase space
representation as in Eq.~(\ref{eq:wignerfunction}). It also
can be used to express the inverse transformation as
\begin{align}
  \hat{A} &= \int d^2x\, \text{Smb}[\hat{A}](x)\, \hat{\Delta}(x).
\label{eq:quantization}\end{align}
 The QM expectation value of an operator $\hat A$ is equal to the phase space average
\begin{align}
  \langle \hat{A} \rangle_{\text{QM}} &= \text{Tr} \hat A \rho =
  \int d^2x\, W(x)\, \text{Smb}[\hat{A}](x)\; .
\label{eq:weylExpec}\end{align}

Two convenient and alternative forms of the quantizer  are
\begin{align}
   \hat{\Delta}(x) &= \frac{ 1}{(2\pi)^2} \int d^2k\,
   e^{ik\cdot(x-\hat{x})}
\label{eq:quantizerFT}\\ &= \frac{ 1}{\pi\hbar} \hat{D}(\alpha_x) \, \hat{\Pi}
  \, \hat{D}^\dagger (\alpha_x)
\label{eq:quantizerAlpha}\end{align}
 (see App.~\ref{app:quantizer}),
where $\hat{x}=(\hat{q},\hat{p})$ are the canonical quantum
observables, $[\hat q, \hat p] = i\hbar$,  and $\hat{\Pi}$ denotes the
parity operator with $ \hat{\Pi}\hat{x} \hat{\Pi} =- \hat{x} $.

The quantity $\hat{D}(\alpha_x)=\exp (\alpha_x \hat{a}^\dagger
-\alpha_x^* \hat{a})$ denotes the  unitary coherent state shift operator with shift amplitude
$\alpha_x = \frac{1}{ \sqrt{2}} (\frac{ q}{L} +i \frac{ L}{\hbar}p)$,
where $L$ is an arbitrary length scale.
Additionally, $\alpha_x$ also represents the Weyl symbol of the
harmonic oscillator  annihilation operator
$\hat{a} = \frac{1}{ \sqrt{2}} (\frac{ \hat{q}}{L} +i \frac{ L}{\hbar}
\hat{p})$.
With this notation, one can consider the quantizer as a function of
the complex variable $\alpha_x$ instead of the phase space variable $x$.
In the following we will drop the index $x$ and work directly
 with the
complex  phase space coordinate $\alpha$.
Then the quantization statement  (\ref{eq:quantization}) can be written as
\begin{align}
  \hat{A} &= \frac{ 2}{\pi} \int d^2\alpha\,
  \text{Smb}[\hat{A}](\alpha)\,
  \hat{D}(\alpha) \, \hat{\Pi}
  \, \hat{D}^\dagger (\alpha),
\label{eq:quantizaion2}\end{align}
with $d^2\alpha=d\text{Re}\alpha\, d\text{Im}\alpha$.

For the special choice $\hat{A}=\hat{{\cal B}}$,
phase space representation (\ref{eq:quantizaion2}) suggests a
decomposition of the Bell operator $\hat{{\cal B}}$
in terms of projectors on coherently shifted
eigenstates of the parity operator,
weighted by the Bell operator's symbol
${\cal B}(\alpha)\equiv \text{Smb}[\hat{{\cal B}}](\alpha)$.
One representation of the
parity operator in terms of
projectors $\hat{P}_n$ on harmonic oscillator eigenstates $|n
\rangle $ is given by
\begin{align}
  \hat{\Pi}= \sum_{n=0}^\infty (-1)^n \hat{P}_n.
\label{eq:parityDecomp}\end{align}

Thus the Bell operator $\hat{{\cal B}}$ has the expansion
\begin{align}
  {\hat {\cal B}} &= \frac{ 2}{\pi} \int d^2\alpha
   \sum_{n=0}^\infty (-1)^n
  {\cal B}(\alpha)
  \hat{D}(\alpha) \,\hat{P}_n
  \, \hat{D}^\dagger (\alpha).
\label{eq:BellOpDecompAlphaN}\end{align}
This is just the form of Eq.~(\ref{eq:BellOpDecomp}) with
$ \int \hspace{-3.3mm} {\scriptstyle  \sum }du
 = \int d^2\alpha  \sum_n$, projectors
\begin{align}
  \hat{P}_n(\alpha) =  \hat{D}(\alpha) \,\hat{P}_n
  \, \hat{D}^\dagger (\alpha)
\label{eq:measurementProjectors}\end{align}
and weights
$w_n(\alpha) = \frac{ 2}{\pi} (-1)^n   {\cal B}(\alpha)$.
Note that for $\alpha \neq \beta$ the projectors $\hat{P}_n(\alpha) , \hat{P}_m(\beta)$
are non-commuting.

The HV bound in Theorem~1 can also be expressed as a function of the
symbol ${\cal B}(\alpha)$.
 It is shown in App.~\ref{app:HVboundSP} that, in this case,
\begin{align}
  \langle \hat{{\cal B}}^2 &  \rangle_\text{HV} =
  \frac{ 4}{\pi^2} \int d^2\alpha \, d^2\alpha'\,
  \sum_{n=0}^\infty
  {\cal B}(\alpha)\,   {\cal B}(\alpha')\,
\nonumber \\ &\hspace{5mm}\times
 \langle n | \hat{D}^\dagger (\alpha) \rho
   \hat{D}(\alpha)
   |n \rangle
     \langle n| \hat{D}(2(\alpha'-\alpha))
    |n \rangle  .
\label{eq:BIupperBoundSP}\end{align}

Because decomposition (\ref{eq:BellOpDecompAlphaN}) can be
implemented with an arbitrary 1D Hermitian operator, the choice
of Bell operator is not restricted.  This conclusion readily extends to Hilbert
spaces, ${\cal H} = L^2(\mathds{R}^N,\mathds{C},d^Nq)$
of wave functions that depend on $N$ spatial variables.
However, the general remarks given at the end of Sec.~\ref{sec:BImain}
still apply: not all expansions (\ref{eq:BellOpDecompAlphaN}) will lead to BI violation. In the next two
sections we provide specific examples of BI violation in phase space.

\section{Single particle Bell violation } \label{sec:loc}
An example of BI violation can now be constructed as follows.
We consider a  single-particle system prepared in state
\begin{align}
   \rho =|1 \rangle \langle1| .
\end{align}
The Wigner function for this state is given by $W(\alpha) =
(\pi\hbar)^{-1} e^{-2|\alpha|^2}(4|\alpha|^2-1)$, which
is negative on the disk $|\alpha|<\frac{ 1}{2}$.
We want to construct a Bell operator that is sensitive to the
negativity of the Wigner function, although this is not
a necessary requirement:
some entangled quantum states do have a positive Wigner function
\cite{FoundPhys36-546}
and positivity of the Wigner function is not sufficient to ensure consistency
with HV models \cite{PhysRevA.79.014104}.
Furthermore, Revzen {\it et al.} \cite{revzen:022103} have shown that a
dichotomic continuous variable BI can be violated with a non-negative
Wigner function.

We define the Bell operator by choosing the Weyl symbol
\begin{align}
  {\cal B}(\alpha)&= 1-2 \theta(1-4|\alpha|^2) ,
\label{eq:symbolLocalExample}\end{align}
where $\theta $ is the step function. This symbol
is equal to the sign of the Wigner function;
  in addition it is just a function of $|\alpha|$. In this
  circumstance Theorem~2 (see App.~\ref{app:BellOpSP}) shows that
the corresponding operator $\hat{\cal{B}}$ is a sum of projectors $|n\rangle\langle n|$ with eigenvalues
\begin{align}
   {\cal B}_n &=
    1 - (-1)^n  \left . \frac{ 1}{n!} \frac{ d^nG}{dt^n} \right |_{t=0}
\label{eq:SPBellOp}\\
   G(t) &=  \frac{2 \left(1- e^{\frac{ 1}{2}\frac{t+1}{t-1}} \right)}{t+1}\,.
\end{align}
A suitable non-commuting operator expansion for this Bell operator
is the phase space representation (\ref{eq:BellOpDecompAlphaN}) with projectors $\hat P_n(\alpha)$.
For this choice of state and Bell operator we have evaluated
Eq.~(\ref{eq:BIupperBoundSP}) and $\langle \hat{{\cal B}} \rangle_{\text{QM}} $
and found that
\begin{align}
  \langle \hat{{\cal B}} \rangle_{\text{QM}} &= \frac{4}{\sqrt{e}}-1 \quad \approx
  1.426
\label{eq:localBIviol0}\\
   \langle \hat{{\cal B}}^2  \rangle_\text{HV} &\approx 1.422
\label{eq:localBIviol}\end{align}
(see App.~\ref{app:localBIviol}). Hence, $|\langle \hat{{\cal B}} \rangle_{\text{QM}}|^2 \approx 2.03 >
 \langle \hat{{\cal B}}^2 \rangle_\text{HV} $,
 so that the generalized BI is violated.

We conclude this section with two remarks. First, it is not required
that the Bell operator decomposition is based on rank-1 projectors.
Instead of using Eq.~(\ref{eq:parityDecomp}) one therefore could
employ a more coarse-grained decomposition of the form
$\hat{\Pi} = \hat{P}_\text{even} -\hat{P}_\text{odd}$, where the two
projectors extract the even and odd part of a spatial wave function,
respectively. However, it is not hard to see that this decomposition
will not lead to BI violation for any choice of $\hat{{\cal B}}$. This
illustrates that BI violation depends not only on the choice of
state $\rho$ and Bell operator $\hat{{\cal B}}$, but also on the way
in which the latter is measured.

Our second remark concerns the negativity of the Wigner function.
The BI derived in this section essentially tests the compatibility of the
Wigner function's negative part with the axioms of deterministic
HV theories. Our result can therefore be interpreted as quantitative evidence
for the ``quantumness'' of a non-positive Wigner function.
This evidence bears some similarity with another measure of
non-classicality for negative Wigner functions \cite{1464-4266-6-10-003,PhysRevA.60.4034}
that has been introduced outside the context of Bell inequalities.
However, we emphasize that this does not imply that positive Wigner
functions are necessarily
classical. Our results only indicate that a Wigner function with negative
values can be in disagreement with deterministic HV theories.

\section{Bi-partite Bell violation}\label{sec:nonloc}
Most experimental tests of Bell inequalities are carried out for bi-partite systems.
So it is of interest to present an example of this type.
The phase space decomposition of a general two-particle
operator takes the form
\begin{align}
  \hat{{\cal B}} &=
     \frac{ 4}{\pi^2}  \int d^2\alpha_1  \int d^2\alpha_2\,
   {\cal B}(\alpha_1,\alpha_2)\,
  \hat{D}(\alpha_1) \,  \hat{\Pi}_1 \hat{D}^\dagger (\alpha_1)
\nonumber \\ &\hspace{4mm}
  \otimes
  \hat{D}(\alpha_2) \,  \hat{\Pi}_2 \hat{D}^\dagger (\alpha_2)
\\ &=
 \frac{ 4}{\pi^2}  \int d^2\alpha_1  \int d^2\alpha_2\,
   {\cal B}(\alpha_1,\alpha_2)\,
 \sum_{n_1,n_2} (-1)^{n_1+n_2}
\nonumber \\ &\hspace{4mm}\times
\hat{P}_{n_1}(\alpha_1) \otimes
  \hat{P}_{n_2}(\alpha_2)
\label{eq:nonlocBellExpan}\end{align}
which is a direct generalization of
Eq.~(\ref{eq:BellOpDecompAlphaN})
\footnote{In the bi-partite case Bell operator
  (\ref{eq:BellOpDecomp}) has some similarity with chained
 Bell inequalities introduced by Braunstein and Caves
 \cite{Braunstein199022}. However, chained BIs are conceptually
 different from the ones considered here.}.
The projectors $\hat{P}_{n}(\alpha) $ correspond to
local observables and are defined
in Eq.~(\ref{eq:measurementProjectors}). In an experiment,
the mean value of $\hat{{\cal B}}$ would be determined by
local measurements of these observables.

In the same fashion as in the previous section we can evaluate
Eq.~(\ref{eq:HVupperBound}) to find
\begin{align}
  \langle \hat{{\cal B}}^2  \rangle_\text{HV} &=
     \frac{ 4}{\pi^2}  \int d^2\alpha_1  \, d^2\alpha_2\,
   {\cal B}(\alpha_1,\alpha_2)\,
  \sum_{n_1,n_2} (-1)^{n_1+n_2}
\nonumber \\ &\hspace{4mm}\times
   \text{Tr}\left ( \rho \hat{P}_{n_1}(\alpha_1) \otimes
  \hat{P}_{n_2}(\alpha_2) \right )
\nonumber \\ &\hspace{4mm}\times
     \text{Tr}\left (  \hat{{\cal B}} \hat{P}_{n_1}(\alpha_1) \otimes
  \hat{P}_{n_2}(\alpha_2) \right ).
\label{eq:heq5}\end{align}

To demonstrate BI violation, we consider two particles prepared in the Bell state
$\rho = |\psi_\text{Bell} \rangle \langle \psi_\text{Bell} |$, with
\begin{align}
 |\psi_\text{Bell} \rangle &= \frac{ 1}{\sqrt{2}} (|0 \rangle  \otimes
  |1 \rangle -|1 \rangle  \otimes
  |0 \rangle ) .
\label{eq:BellSTate}\end{align}
The Wigner function of this state takes the form
\begin{align}
  W_\text{Bell}(\alpha_1,\alpha_2) &= \frac{ 1}{\pi^2\hbar^2} e^{-|\alpha_1|^2-|\alpha_2|^2}
  \left (2 |\alpha_1-\alpha_2|^2 - 1 \right )\; .
\label{eq:WignerBell}\end{align}
This corresponds to a product $W_{00}(\alpha_1+\alpha_2) W_{11}(
\alpha_1-\alpha_2)$ of the ground state in the center-of-mass coordinates and first excited state in
the relative coordinates. To test the negativity in relative
coordinates, we chose the symbol of the Bell operator as
\begin{align}
  {\cal B}(\alpha_1,\alpha_2)&=
  1-2\theta(1-2|\alpha_1-\alpha_2|^2).
\end{align}
Because of the product structure of the Wigner function and the fact
that the Bell operator only tests the relative coordinate
$\alpha_1-\alpha_2$, the result for the QM mean value is the same
as in the single-particle case and given by Eq.~(\ref{eq:localBIviol0}).
The evaluation of the upper bound (\ref{eq:heq5})
is presented in App.~\ref{app:BellStateExample} and yields
$ \langle \hat{{\cal B}}^2 \rangle_\text{HV}
\approx 1.27$. Hence, we have again a BI violation
$|\langle \hat{{\cal B}} \rangle_{\text{QM}}|^2 \approx 2.03 >
 \langle \hat{{\cal B}}^2  \rangle_\text{HV} $.

It is interesting to note that the degree of violation in this
non-local example is larger than in the single-particle case.
We believe that the reason for this is that the restriction of the
projection operators to be local effectively increases the number
of projection measurements needed to determine $\langle \hat{{\cal B}}
\rangle_{\text{QM}}$. If non-local measurements were possible, we could have
decomposed the Bell operator as in Eq.~(\ref{eq:BellOpDecompAlphaN}),
with $\hat{P}_n$ and $\hat{D}(\alpha)$ acting on the relative coordinate
between the two particles. This smaller decomposition would have
produced the same HV bound as the single-particle example.
We conjecture that generally a more fine-grained decomposition of a given
Bell operator $\hat{{\cal B}}$ may lead to a lower HV bound.

\section{Conclusion}
We have proposed generalized Bell inequalities, which are
constructed by decomposing a general Bell operator $\hat{{\cal B}}$
into a set of non-commuting projection operators. The derivation
of these inequalities is based on Gleason's theorem, so that a
violation of it would rule out a non-contextual hidden-variable
interpretation of quantum physics.

We have shown that the CHSH inequality may be considered as a special
case of the generalized BI and presented two examples of BI violation
in quantum phase space. The examples test the negativity of the
Wigner function for a single particle and for a two-particle system.
A larger degree of BI violation is obtained in the second example,
for which all measurement observables have a product structure
similar to the CHSH inequality.

The proposed inequalities may be applied to different combinations
of Bell operators and quantum states, or to different decompositions
of a given Bell operator. There is no restriction on the size or
partition of the quantum system, except that the dimension of Hilbert
space must be larger than 2. This opens the possibility to search for
generalized BI violations under very general circumstances,
including the natural decomposition of a general Bell operator
in terms of its Weyl symbol presented in Sec.~\ref{sec:quant}.

There are also several formal aspects of the proposed BI that would
be of interest. The most interesting question is probably whether
Eq.~(\ref{eq:genBI}) could be derived without using
Gleason's theorem, so that a broader class of HV theories could
be ruled out if a generalized BI violation is experimentally confirmed.
That this is possible at least in special cases is demonstrated by the
example of the CHSH inequality. We have shown that it may be derived
using  Eq.~(\ref{eq:genBI}), but it is well known that there are other
ways to prove it that do not rely on Gleason's theorem \cite{Redhead:Incompleteness}.
Thus, it is conceivable that Eq.~(\ref{eq:genBI})  may serve as a tool to identify
BIs that also rule out contextual local HV theories.

Other extensions of our proposal include the question for
which decomposition of a given Bell operator the HV bound is
minimized, or to find examples of BI violation that can be
realized with specific experimental setups. This may be the
topic of further studies.

\acknowledgments
This project was funded by NSERC and ACEnet.
T.~A.~O. is grateful for an appointment as James Chair, and
K.-P.~M. for a UCR grant from St.~Francis Xavier University.

\begin{appendix}
\section{ The Collapse BI}\label{app:proofBI}
In this section we
provide the axiomatic foundations of hidden variable theories, construct a proof of Theorem 1 and discuss how determinism is implemented.

The first stage of our proof utilizes the method of Cavalcanti
{\em et al}  \cite{PhysRevLett.99.210405}, which
is based on the fact
that for a random variable $\bar{{\cal B}}$ in an HV theory
the following variance inequality  must hold,
\begin{align}
   |\langle \bar{{\cal B}} \rangle_\text{HV} |^2 &\leq \langle \bar{{\cal B}}^2
    \rangle_\text{HV}  \; .
\label{eq:varIneq}\end{align}

The HV framework employed here is that established in the works of Fine and Malley, specifically the axioms HV(a-d) as formulated in Malley \cite{PhysRevA.58.812}.
Hidden variable theories link a family  of quantum observables ${\cal O= \cal O}(\widehat H,\Xi,\rho)$ with a
corresponding family of HV observables, $\Omega=\Omega(\Lambda, {\cal F},\mu )$. The quantum system density matrix is $\rho$.  Here $\Xi$ is a subset of observables on Hilbert space  ${\cal H}$, which includes all those that appear in a Bell inequality of interest.  In the case of Theorem 1, this operator collection would include the projectors and their products appearing in expansion (\ref{eq:BellOpDecomp}) of $\widehat B$. \smallskip

The triplet $\Omega(\Lambda, {\cal F},\mu )$ represents a  classical probability space
wherein variable $\lambda \in \Lambda$ is the HV state of the system and $\Lambda$ is the set of all ``complete state specifications'' \cite{fine1982hidden};
${\cal F}$ is a (Borel) $\sigma$-algebra of subsets
of $\Lambda$; and, $\mu: {\cal F}\rightarrow [0,1]$ is a unit normalized probability measure. In the hidden variable picture an allowed observable, say  $\widehat A\in \Xi$,  is represented by a $\mu$-measurable function (random variable) $A(\lambda)$ and expectation values result from an integral over $\Lambda$ weighted with measure $\mu$.

In a {\it deterministic} model the (unique) values of $A(\lambda)$ are those fixed by a measurement of $A$ in the HV state $\lambda$. In full detail, the model is defined by the following four axioms;  in each, it is required that the quantum observables are restricted to the set $\Xi$.
\\[1mm]
 HV(a) (the spectrum rule) restricts possible values
of a random variable to the spectrum of the respective operator in
quantum theory.
\\[1mm]
HV(b) (the sum rule) requires additivity of the values of two random
variables that correspond to commuting operators.
\\[1mm]
 HV(c) (the first-order margins rule) states that the marginal probabilities
agree with QM, $\langle E_u \rangle_\text{HV} =  \text{Tr}(\rho \hat{P}(u))$.
\\[1mm]
HV(d)  (the second-order margins rule) requires that
$ \langle E_u E_v \rangle_\text{HV} = \text{Tr}(\rho \hat{P}(u) \hat{P}(v))$
if the projectors $\hat{P}(u) , \hat{P}(v)$ commute.
\\[1mm]
Of  particular interest are random variables
$E_u$ that correspond to quantum projectors $\hat{P}(u)$.
Rule HV(a) ensures that the equivalent observable to a projector $\hat{P}(u)$
in an HV theory must correspond to a random variable of the form
\begin{align}
  E_u(\lambda) = \left \{
  \begin{array}{cc}
           1  &\quad \lambda \in S(u) \\ 0 & \text{otherwise}
           \end{array} \right . ,
\end{align}
where hidden variable $\lambda \in\Xi$ and $S(u)$ is the set of
all outcomes for which $E_u(\lambda)=1$.
The expectation value of this random variable is
equal to the probability to find a unity value,
\begin{align}
  \langle E_u \rangle_\text{HV} &= \mu(S(u))=
  \int E_u(\lambda)d\mu(\lambda)
\\ & =  \int E_u(\lambda)\, \mu'(\lambda) \, d\lambda .
\end{align}
The last relation provides a link between the more abstract notion
of HV theories as a classical probability space and Bell's original
notation. Hidden variables $\lambda$ can be considered as a
parametrization of the sample space and $\mu'(\lambda) $
provides the probability density with respect to this parametrization.

The first step in our proof is to define a new classical random variable by superposing the $E_u$ to match the form of expansion (\ref{eq:BellOpDecomp})
\begin{align}
  \bar{{\cal B}}(\lambda) &= \int \hspace{-4mm} {\textstyle \sum} du\,  w(u)\, E_u(\lambda)  .
\label{eq:BellOpDecompHV} \end{align}
 The  HV mean value of $ \bar{{\cal B}}(\lambda) $ is given by
\begin{align}
  \langle \bar{ {\cal B}} \rangle_\text{HV} &=
  \int \hspace{-4mm} {\textstyle \sum} du\,  w(u)\, \langle E_u \rangle_\text{HV}.
\label{eq:BellOpDecompHVmean} \end{align}

Using axiom {HV\,\!(c)} in Eq.~(\ref{eq:BellOpDecompHVmean}) we obtain
$ \langle  \bar{{\cal B}} \rangle_\text{HV} =  \langle  \hat{{\cal B}}
\rangle_\text{QM}$, i.e., the mean value of the classical
random variable $\bar{{\cal B}}(\lambda)$ should agree with that
of the QM Bell operator (\ref{eq:BellOpDecomp}).
This implies, as a consequence of (\ref{eq:varIneq}),   that for QM to be compatible with
the axioms of HV theories,
$|\langle  \hat{{\cal B}} \rangle_\text{QM}|^2$ should be bounded
by the HV expectation value
\begin{align}
  \langle \bar{{\cal B}}^2 \rangle_\text{HV} &=
 \int \hspace{-4mm} {\textstyle \sum} du\, dv\,
   w(u) \, w(v)\,  \langle E_u E_v \rangle_\text{HV}.
\end{align}

The joint probability to find the value 1 in both observables,
$\langle E_u E_v \rangle_\text{HV} = \mu(S(u)\cap S(v))$, can be
expressed in terms of classical conditional probabilities
$\mu(A|B) = \mu(A\cap B)/\mu(B)$, as
\begin{align}
 \langle E_u E_v \rangle_\text{HV}  &= \mu(S(u)|S(v))\, \mu(S(v))
\\ &= \mu(S(v)|S(u)) \, \mu(S(u)),
\end{align}
so that
\begin{align}
  \langle  \bar{{\cal B}^2}  \rangle_\text{HV}
&=
 \int \hspace{-4mm} {\textstyle \sum} du\, dv\,
   w(u) \, w(v)\, \mu( S(v) | S(u)) \, \mu(S(u)).
\end{align}

The analog of conditional probabilities in quantum theory
is the  L\"uders rule \cite{Bobo2010}
$\text{Tr}(\rho_{u}\hat{P}(v))$.
This represents the probability to measure
$\hat{P}(v)$ under the condition that $\hat{P}(u)$ has been
measured before.
It has been shown that the L\"uders rule
is the unique extension of classical conditional
probabilities to quantum mechanics (see p.~288 of Ref.~\cite{Beltrametti:QM81}).
The key ingredient in our proof is
Malley's result  \cite{PhysRevA.58.812}  that quantum and classical HV conditional
probabilities must agree for a pair of not necessarily commuting projectors,
\begin{align}
   \mu( S(v) | S(u)) &=
  \text{Tr}(\rho_{u}\hat{P}(v)).
\label{eq:condProbRule} \end{align}

The proof of Eq.~(\ref{eq:condProbRule}) is based on the hidden variable model HV(a--d) and Gleason's theorem \cite{gleason1957measures}. The latter restricts the dimension of Hilbert space to ${\dim(\cal{H})} \geq 3 $.
Employing (\ref{eq:condProbRule}) we can express the HV upper bound as
\begin{align}
  \langle  \bar{{\cal B}}^2\rangle_\text{HV} &=
 \int \hspace{-4mm} {\textstyle \sum} du\, dv\,
   w(u) \, w(v)\,  \text{Tr}(\rho_{u}\hat{P}(v))
  \, \mu(S(u))
\\ &=
   \int \hspace{-4mm} {\textstyle \sum} du\, dv\,
   w(u) \, w(v)\,  \text{Tr}(\rho_{u}\hat{P}(v))
  \, \text{Tr}(\rho\hat{P}(u)).
\end{align}
Combined with Eq.~(\ref{eq:varIneq}) this is the statement of Theorem 1.
We remark that we have changed the notation from
$ \langle  \bar{{\cal B}}^2\rangle_\text{HV} $ to
$ \langle  \hat{{\cal B}}^2 \rangle_\text{HV} $
because the HV bound can be expressed through properties
of QM operators alone.

The quantum upper bound of Eq.~(\ref{eq:varIneq}) can be
related to the HV bound in the following way
\begin{align}
  \langle \hat{{\cal B}}^2  \rangle_\text{QM} &=
  \int \hspace{-4mm} {\textstyle \sum} du\, dv\,
   w(u) \, w(v)\,   \text{Tr}(\rho \hat{P}(u) \hat{P}(v))
\\ &=
  \int \hspace{-4mm} {\textstyle \sum} du\, dv\,
   w(u) \, w(v)\,   \text{Tr}(\rho \hat{P}^2(u) \hat{P}(v))
\\ &=
  \int \hspace{-4mm} {\textstyle \sum} du\, dv\,
   w(u) \, w(v)\,
    \Big ( \text{Tr}(\rho \hat{P}(u) \hat{P}(v) \hat{P}(u) )
\nonumber \\ & \hspace{1cm}
    + \text{Tr}(\rho \hat{P}(u) [\hat{P}(u), \hat{P}(v)] )
  \Big ).
\end{align}
Recognizing that the first term in parentheses reproduces the HV
upper bound this leads to Eq.~(\ref{eq:boundsDifference}).

In the remainder of this appendix we will clarify some of the basic
features of deterministic HV models as established by Fine and Malley.

First note that it is the spectrum rule HV(a)
that implements determinism in HV theories. Axiom HV(a) implies that all
observables in an HV model assume specific values. These values are
not necessarily known to us, and this uncertainty about their value is captured in the
measure density distribution $\mu'(\lambda)$; but we do know that one of
these values would be assumed in any run of the experiment. This is in
contrast to quantum mechanics where it cannot be said that an
observable $\widehat{A}$ assumes a specific (spectral) value unless the system is
prepared in (an incoherent mixture, but not a superposition of
\cite{jmp/37/6/10.1063/1.531530}) eigenstates of $\widehat {A}$. In a deterministic
HV theory it is in principle possible to prepare the system in
a state where all properties of all observables the system are simultaneously and exactly known.
In the literature, such states have been called dispersion-free states
\cite{Hemmick2012} or completed states \cite{Beltrametti:QM81}. Such
states would be described by a
distribution $\mu'(\lambda)$ that takes the form of a
Dirac distribution. However a deterministic HV model does not restrict
the form of $\mu'(\lambda)$.
 In a measurement where the marginal rules HV(c) and HV(d) apply, the resulting $\mu'(\lambda)$ will generally be a distribution with dispersion. 

Although the goal of hidden variable theories is to be as consistent as possible with the predictions of quantum mechanics they nevertheless
differ in a number of ways.   Key among these is that the HV(a-d) framework provides a
joint distribution (probability measure, $\mu$) that applies to all
pairs of projectors in 
${\Xi}$.  In detail,
the HV values obey $\langle E_{u} E_{v} \rangle_{HV} = \langle E_{v} E_{u}  \rangle_{HV}, \ \forall\, \hat P(u), \hat P(v) \in \Xi$.  In QM
it is known \cite{gudder1979stochastic} [Thm. 2.1] that a joint probability distribution for a set of non-commuting operators, such as ${\Xi}$, cannot exist.  Also some combinations of the axioms lead to new useful identities, for  example it can be shown that, given HV (a), the product rule is equivalent to HV (b),
see rule HV (b$_1$) of Ref.~\cite{PhysRevA.58.812}. 

Deterministic HV theories also differ from contextual HV
theories. To explain the difference we consider
the joint probability distribution to measure the eigenvalues
$a,b$ of two observables $\hat{A},\hat{B}$, which in a deterministic HV
is given by
$
  \mu(a,b) = \int \mu'(a,b,\lambda)\, d\lambda
$. On the other hand, in a contextual HV theory
the probability density
$\mu'(a,b,\lambda) = \mu'(a,b,\lambda | \hat{A},\hat{B}) $ may depend on the
measurement settings to detect observables $\hat{A}, \hat{B}$.
If the system is set up to measure observable $\hat{B}'$ instead of
$\hat{B}$, the probability density
$ \mu'(a,b,\lambda | \hat{A},\hat{B}') $ to find eigenvalues $a,b$
for the observables $\hat{A},\hat{B}$ may be different than for the
original setting.  One may say that in the contextual
setting the HV model has more than one probability measure (one for each
experimental setup).

The work of Fine \cite{fine1982joint} establishes that, in the context
of bi-partite systems and the original BI,
axioms HV(a-d) are equivalent to the HV model conditions
assumed by Bell \cite {bell1966problem} and Kochen-Specker
\cite{Koc}. Specifically, he proved that the necessary and
sufficient condition for the existence of a deterministic HV model  is that
the original BI (\ref{eq1}) is not violated (Proposition 2 in Ref.~\cite{fine1982hidden}).

While factorizablility is
widely accepted for two observables that belong to two space-like
separated systems, this is not the case for observables of an
individual system. Because Gleason's theorem does not address
contextuality \cite{Hemmick2012}, and because the upper bound
(\ref{eq:HVupperBound}) contains products of observables
that belong to a single subsystem, an experimental violation of our
`collapse' BI  would exclude only
deterministic HV models but not contextual HV theories.

\section{ Quantizer forms  }\label{app:quantizer}
\noindent Proof of Eq.~(\ref{eq:quantizerFT}): Let
$\hat{\delta} \equiv \int \frac{ d^2k}{(2\pi)^2}  e^{ik\cdot(x-\hat{x})}$. Then
\begin{align}
\hat{\delta} &=
  \int \frac{ d^2k}{(2\pi)^2}  e^{ik\cdot x}
  e^{-i k_q \hat{q}}  e^{-i k_p \hat{p}} e^{\frac{ 1}{2} k_q k_p i
    \hbar}
\\ &=
   \int \frac{ d^2k\, dq'}{(2\pi)^2}  e^{ik\cdot x}
  e^{-i k_q q'} |q' \rangle \langle q'-\hbar k_p|  e^{\frac{ 1}{2} k_q k_p i
    \hbar}
\\ &=
  \int \frac{ dq'\,  dk_p}{2\pi}
  e^{ik_p p}
   |q' \rangle \langle q'-\hbar k_p|
  \delta(q-q' +\frac{ \hbar}{2} k_p )
\\ &=
    \frac{ 1}{2\pi\hbar}
  \int dq'\, e^{-i p q'/\hbar}
   |q -\frac{1}{2} q' \rangle \langle q +\frac{ 1}{2} q'|
  \quad \Box
\end{align}
Proof of Eq.~(\ref{eq:quantizerAlpha}):
We start by observing that
$\hat{D}(\alpha_x)=\exp\left (\frac{ i}{\hbar}
 (\hat{q}p-\hat{p}q)\right)$. Hence,
\begin{align}
 \hat{D}(\alpha_x) & \frac{ \hat{\Pi}}{\pi\hbar}
  \, \hat{D}^\dagger (\alpha_x)
  = \int \frac{ dq'}{\pi\hbar}
   \hat{D}(\alpha_x)|q' \rangle \langle -q'| \hat{D}^\dagger
   (\alpha_x)
\\ &=  \int \frac{ dq'}{\pi\hbar}\int dp
  e^{\frac{ 2i}{\hbar}q' p}
  e^{-\frac{ i}{\hbar} q\hat{p}}
  |q' \rangle \langle -q'|
  e^{\frac{ i}{\hbar} q\hat{p}}
\\ &=
    \int \frac{ dq'}{\pi\hbar}
  e^{\frac{ 2i}{\hbar}q' p}
  |q+q' \rangle \langle q-q'| \quad \Box
\end{align}

\section{Bell bound $\langle \hat{{\cal
      B}}^2\rangle_{HV}$}\label{app:HVboundSP}
For decomposition (\ref{eq:BellOpDecompAlphaN}),
HV bound (\ref{eq:HVupperBound}) takes the form
\begin{align}
  \langle \hat{{\cal B}}^2 &  \rangle_\text{HV} =
  \frac{ 2}{\pi} \int  d^2\alpha \sum_n
  {\cal B}(\alpha)\, (-1)^n
\nonumber \\ &\hspace{4mm} \times
  \text{Tr}\left (  \rho
   \hat{D}(\alpha) \,  \hat{P}_n
  \, \hat{D}^\dagger (\alpha)
   \hat{{\cal B}}
   \hat{D}(\alpha) \,  \hat{P}_n
  \, \hat{D}^\dagger (\alpha)
  \right )
\\ &=
  \frac{ 2}{\pi}\int d^2\alpha \sum_n
  {\cal B}(\alpha)\, (-1)^n
  \text{Tr}\left (  \rho
   \hat{D}(\alpha) \,  \hat{P}_n
  \, \hat{D}^\dagger (\alpha)
  \right )
\nonumber \\ &\hspace{4mm}\times
   \text{Tr}\left (
   \hat{{\cal B}}
   \hat{D}(\alpha) \,  \hat{P}_n
  \, \hat{D}^\dagger (\alpha)
  \right ).
\label{eq:ex3-heq1}\end{align}
Using the fact that $\hat{P}_n$ is a rank 1 projector we evaluate the second trace
\begin{align}
   \text{Tr} & \left (
   \hat{{\cal B}}
   \hat{D}(\alpha) \,  \hat{P}_n
  \, \hat{D}^\dagger (\alpha)
  \right ) =
   \frac{ 2}{\pi}\int  d^2\alpha' \sum_m
  {\cal B}(\alpha')\, (-1)^m
\nonumber \\ &\hspace{4mm}\times
     \text{Tr}\left (
  \hat{D}(\alpha') \,  \hat{P}_m
  \, \hat{D}^\dagger (\alpha')
   \hat{D}(\alpha) \,  \hat{P}_n
  \, \hat{D}^\dagger (\alpha)
  \right )
\\ &=
  \frac{ 2}{\pi}\int  d^2\alpha'\,
  {\cal B}(\alpha')\,
     \text{Tr}\Big (
  \hat{D}(\alpha') \,  \hat{\Pi}\, \hat{D}^\dagger (\alpha')
\nonumber \\ &\hspace{3cm}\times
 \hat{D} (\alpha)
    \hat{P}_n
  \, \hat{D}^\dagger (\alpha)
  \Big )
\\ &=
   \frac{ 2}{\pi} \int d^2\alpha'\,
  {\cal B}(\alpha')\,
     \text{Tr}\!\left (
  \hat{\Pi}
  \, \hat{D}^\dagger (\alpha'\!\!-\!\alpha)
   \hat{P}_n \hat{D}(\alpha'\!\!-\!\alpha)
  \right )\,.
\end{align}
The parity and the shift operator have the commutation
relation
\begin{align}
   \hat{\Pi} \hat{D}(\beta) &= \hat{D}(-\beta)  \hat{\Pi}\,.
\label{eq:parityOpComm}\end{align}
Furthermore, $\hat{\Pi}\hat{P}_n = (-1)^n \hat{P}_n$ holds.
We thus get for the trace
\begin{align}
   \text{Tr}&\left (
   \hat{{\cal B}}
   \hat{D}(\alpha) \,  \hat{P}_n
  \, \hat{D}^\dagger (\alpha)
  \right ) =
\nonumber \\ &\hspace{3mm}
   \frac{ 2}{\pi} \int d^2\alpha'\,
  {\cal B}(\alpha') (-1)^n
     \text{Tr}\left (
   \hat{P}_n \hat{D}(2(\alpha'-\alpha))
  \right ),
\end{align}
so that
\begin{align}
  \langle \hat{{\cal B}}^2 & \rangle_\text{HV} =
   \frac{ 4}{\pi^2} \int d^2\alpha \, d^2\alpha'\,
  \sum_n
  {\cal B}(\alpha)\,   {\cal B}(\alpha')\,
 \label{eq:ex3-heq2} \\ &\hspace{4mm}\times
  \text{Tr}\left (  \rho
   \hat{D}(\alpha) \,  \hat{P}_n
  \, \hat{D}^\dagger (\alpha)
  \right )
     \text{Tr}\left (
   \hat{P}_n \hat{D}(2(\alpha'-\alpha))
  \right ) \quad \Box \nonumber
\end{align}

\section{Bell eigenvalues}\label{app:BellOpSP}
The Bell symbol (\ref{eq:symbolLocalExample}) in the single particle example is just a function of $|\alpha|$.  This structure simplifies the computation of its quantum expectation value as follows.
\\[3mm]
{\bf Theorem 2}: Let $\hat A$ be an operator
with Weyl symbol  $A(\alpha)$.  If $A(\alpha)$ only depends
on $|\alpha|$ then $\hat A$ has the spectral expansion
\begin{equation}\label{thm2a}
    \hat A = \sum_0^\infty \lambda_n \hat P_n
\end{equation}
with eigenvalues
\begin{equation}\label{thm2b}
    \lambda_n = \int_0^\infty d|\alpha|^2 \, A(|\alpha|) 2 (-1)^n
    L_n(4|\alpha|^2) e^{-2|\alpha|^2}\, \, ,
\end{equation}
where $L_n$ is the Laguerre polynomial of order $n$. \\
{\em Proof}: Consider the commutator of $\hat A$ and $\hat P_n$
\begin{align}
  [\hat{ A},\hat{P}_n] &= \frac{ 2}{\pi}\int d^2\alpha\,
  {  A}(|\alpha|)\,
  [\hat{D}(\alpha) \, \hat{\Pi}
  \, \hat{D}^\dagger (\alpha),\hat{P}_n]
\\ &=
   \frac{ 2}{\pi}\int d^2\alpha\,
  {A}(|\alpha|)\,
  [\hat{D}(2\alpha) \, \hat{\Pi} ,\hat{P}_n]\,.
\end{align}
We now make a change of variables $\alpha = r e^{i\phi}$, and carry
out the integration over $\phi$.
Using the Baker-Campell-Hausdorff formula one then finds
\begin{align}
  \int_0^{2\pi} d\phi\, & \hat{D}(2r e^{i\phi})
  =   \int_0^{2\pi} \frac{ d\phi}{2\pi}
  e^{-2r^2} e^{2r e^{i\phi} \hat{a}^\dagger }
   e^{-2r e^{-i\phi} \hat{a} }
\\ &=
  e^{-2r^2} \int_0^{2\pi} d\phi\,
  \sum_{n,m=0}^\infty
  (2r \hat{a}^\dagger )^n
   (-2r \hat{a} )^m e^{i\phi(n-m)}
\\ &=
  2\pi e^{-2r^2}
  \sum_{m=0}^\infty
  (-4r^2)^m  (\hat{a}^\dagger)^m \hat{a}^m.
\end{align}
This operator obviously commutes with $\hat{P}_n$ and $\hat \Pi$ and so
(\ref{thm2a}) is established.

The eigenvalue is given by $\langle n|\hat A|n \rangle$,
which can be evaluated using Eq.~(\ref{eq:weylExpec}). The
symbol of $\hat P_n$ has been derived in
Ref.~\cite{PhysRevLett.83.3758}
and is given by
\begin{equation}\label{Pn}
    P_n(|\alpha|)) = (-1)^n 2 L_n(4|\alpha|^2) e^{-2|\alpha|^2}
\end{equation}
Inserting this into  Eq.~(\ref{eq:weylExpec}) leads to Eq.~(\ref{thm2b})  $\Box$

Note that if $A(\alpha)$ is not real then $\hat A^\dag \neq \hat A$ and the eigenvalues  $\lambda_n$ are complex.

Consider the single particle operator $\hat{{\cal B}}$ with symbol (\ref{eq:symbolLocalExample}). Let ${\cal B}_n$ denote its eigenvalues in the harmonic oscillator expansion of Theorem~2. Then
\begin{align}
   {\cal B}_n &=
  \int d^2|\alpha|\,P_n(|\alpha|)) \,
(1-2 \theta(1-4|\alpha|^2))
\\ &=
  1 -4  \int_{|\alpha|< \frac{ 1}{2}} d^2|\alpha| \, (-1)^n  L_n(4|\alpha|^2) e^{-2|\alpha|^2} \,.
\end{align}
Using again a polar decomposition $\alpha =\frac{ 1}{2}\sqrt{u} e^{i\phi}$ we find
\begin{align}\label{eq:D11}
   {\cal B}_n &=
    1- (-1)^n\int_{0}^{1} du\,
   e^{-\frac{ u}{2}}L_n(u)
\end{align}
By replacing the Laguerre polynomials with their
generating function (Eq.~(22.9.15) of Ref.~\cite{abramowitz}),
we can express the eigenvalues as in Eq.~(\ref{eq:SPBellOp}).

\section{Single-particle BI}\label{app:localBIviol}

The quantum expectation value $\langle \hat{{\cal B}} \rangle_{QM}$ in state
$|1 \rangle \langle 1|$ is just the eigenvalue
${\cal B}_1$ of (\ref{eq:D11}).

To derive Eq.~(\ref{eq:localBIviol}) we use \cite{PhysRev.177.1857}
\begin{align}
    \langle n+a | \hat{D}(\alpha) |n \rangle &=
   e^{-\frac{ 1}{2}|\alpha|^2}\alpha^a \sqrt{\frac{ n!}{(n+a)!}}
   L_n^{(a)}(|\alpha|^2)
\label{eq:shiftOpMatrixElements}\\
   \langle n | \hat{D}(\alpha) |n+a \rangle &=
    \langle n+a | \hat{D}(-\alpha) |n \rangle^*\,.
\end{align}
Equation~(\ref{eq:BIupperBoundSP}) then becomes
\begin{align}
  \langle \hat{{\cal B}}^2 &   \rangle_\text{HV} =
  \frac{ 4}{\pi^2} \int d^2\alpha \, d^2\alpha'\,
  \sum_{n=0}^\infty
  {\cal B}(\alpha)\,   {\cal B}(\alpha')\,
\nonumber \\ &\hspace{5mm}\times
 |\langle n | \hat{D}^\dagger (\alpha) |1 \rangle |^2
     e^{-2|\alpha-\alpha'|^2}
   L_n(4|\alpha-\alpha'|^2)
\\ &=  \frac{ 4}{\pi^2} \int d^2\alpha \, d^2\alpha'\,
  \sum_{n=0}^\infty
  {\cal B}(\alpha)\,   {\cal B}(\alpha')\,
 e^{-2|\alpha-\alpha'|^2}
e^{-|\alpha|^2}
\nonumber \\ &\hspace{5mm}\times
  \frac{ 1}{n!}
   |\alpha|^{2(n-1)}
   (n-|\alpha|^2)^2
   L_n(4|\alpha-\alpha'|^2).
\end{align}
The sum can be simplified using
Eq.~(8.975-3) of Ref.~\cite{GradshteynRyzhik},
\begin{align}
  S(x,y) &\equiv \sum_{n=0}^\infty \frac{ y^n}{n!} L_n(x) \quad =
  J_0(2\sqrt{xy}) e^y,
\label{eq:LaguerreSumRule}\end{align}
where $J_n(x)$ denotes the Bessel function of integer order.

Setting $x=4|\alpha-\alpha'|^2$ and $y=|\alpha|^2$ we have to evaluate
the sum
\begin{align}
    \sum_{n=0}^\infty \frac{y^{n-1} }{n!}
      (n-y)^2   L_n(x) & =
  y \big (
  S(x,y)-2 \partial_y S(x,y)
\nonumber \\ &\hspace{2mm}
  + \partial_y^2S(x,y)+ \frac{ 1}{y} \partial_y S(x,y)
  \big )
\\ &=
  e^y (1-x) J_0(2\sqrt{xy}).
\end{align}
Hence,
\begin{align}
  \langle \hat{{\cal B}}^2   \rangle_\text{HV}
 & = \int d^2\alpha \, d^2\alpha'\,
  {\cal B}(\alpha)\,   {\cal B}(\alpha')\,     f(\alpha,\alpha')
\\
   f(\alpha,\alpha') &\equiv
   \frac{ 4}{\pi^2}
  \frac{ 1-4|\alpha-\alpha'|^2}{
   e^{2|\alpha-\alpha'|^2}}
   J_0(4|\alpha-\alpha'|\, |\alpha|).
\end{align}
The two Weyl symbols
$ {\cal B}(\alpha),   {\cal B}(\alpha') $ are both of the form
(\ref{eq:symbolLocalExample}), which leads to a natural separation
of the integral into four parts. Setting
\begin{align}
  I(R,R') &\equiv  \int_{|\alpha|<R} d^2\alpha \int_{|\alpha'|<R'}
   d^2\alpha'\,   f(\alpha,\alpha'),
\end{align}
we have
\begin{align}
  \langle \hat{{\cal B}}^2   \rangle_\text{HV}
 & =
  I(\infty,\infty)
  -2  I\left ( \frac{ 1}{2} ,\infty\right )
\nonumber \\ &\hspace{4mm}
  -2  I\left (\infty, \frac{ 1}{2} \right )
  +4  I\left (\frac{ 1}{2} , \frac{ 1}{2} \right ).
\end{align}
The first two integrals can be found analytically and have the values
$I(\infty,\infty)=1$ and $ I\left ( \frac{ 1}{2} ,\infty\right )=1-2
e^{-\frac{ 1}{2}}$. We have numerically evaluated the remaining parts
and found $ I\left (\infty, \frac{ 1}{2} \right )\approx 0.0184$
and $ I\left (\frac{ 1}{2} , \frac{ 1}{2} \right )\approx 0.0082$, so
that
$ \langle \hat{{\cal B}}^2  \rangle_\text{HV}
\approx 1.422$.

\section{Bi-partite example}\label{app:BellStateExample}
We start by evaluating the traces appearing in Eq.~(\ref{eq:heq5}).
Using Eqs.~(\ref{eq:shiftOpMatrixElements}) and
(\ref{eq:nonlocBellExpan})
it is straightforward to transform the second trace into
\begin{align}
       \text{Tr} & \left (  \hat{{\cal B}}  \hat{P}_{n_1}(\alpha_1) \otimes
  \hat{P}_{n_2}(\alpha_2) \right ) =
   \frac{4 }{\pi^2}  \int d^2\beta_1 \int d^2\beta_2
\nonumber \\ & \times
   {\cal B}(\beta_1,\beta_2)\,
  e^{-2|\beta_1-\alpha_1|^2} L_{n_1}(4|\beta_1-\alpha_1|^2)
\nonumber \\ &\times
  e^{-2|\beta_2-\alpha_2|^2} L_{n_2}(4|\beta_2-\alpha_2|^2)
  (-1)^{n_1+n_2}.
\end{align}
For state (\ref{eq:BellSTate}), the first trace in
Eq.~(\ref{eq:heq5}) becomes
\begin{align}
   \text{Tr} \big ( \rho & \hat{P}_{n_1}(\alpha_1) \otimes
  \hat{P}_{n_2}(\alpha_2) \big )
\nonumber \\ &=
  \left | \left (\langle n_1|\hat{D}^\dagger (\alpha_1)\otimes
   \langle n_2|\hat{D}^\dagger (\alpha_2) \right ) |\psi_\text{Bell} \rangle
  \right |^2
\\ &=
  \frac{ 1}{2} \Big |
    \langle n_1|\hat{D}^\dagger (\alpha_1)|0 \rangle \,
     \langle n_2|\hat{D}^\dagger (\alpha_2) |1 \rangle
\nonumber \\ & \hspace{5mm}
   -
    \langle n_1|\hat{D}^\dagger (\alpha_1)|1 \rangle \,
     \langle n_2|\hat{D}^\dagger (\alpha_2) |0 \rangle
  \Big |^2
\\ &=
   \frac{ |\alpha_2|^{2(n_2-1)}
     |\alpha_1|^{2(n_1-1)} }{2(n_1)! (n_2)!}
   e^{-|\alpha_1|^2-|\alpha_2|^2}
\nonumber \\ &\hspace{5mm}\times
   \left |
   \alpha_1 (n_2-|\alpha_2|^2)  -
    \alpha_2 (n_1-|\alpha_1|^2)
  \right |^2 .
\end{align}
With this expression, the sum over $n_1$ and $n_2$ in Eq.~(\ref{eq:heq5})
can again be evaluated using relation (\ref{eq:LaguerreSumRule}).
Changing the integration variables from $\beta_i$ to $\gamma_i =
\beta_i-\alpha_i$, i=1,2, yields
\begin{align}
  \langle \hat{{\cal B}}^2 & \rangle_\text{HV} =
  \frac{ 16}{\pi^4}
  \int d^2\alpha_1  \, d^2\alpha_2\,  d^2\gamma_1  \, d^2\gamma_2\,
   e^{-2 \left( | \gamma _1 | ^2+ | \gamma_2 | ^2\right)}
\nonumber \\ & \times
   \Big (
  \left (1 -2 | \gamma_1|^2 -2| \gamma _2|^2\right)
    J_0(4 | \alpha _1 \gamma _1|) J_0(4 |\alpha_2 \gamma _2|)
\nonumber \\ &
  -2 | \gamma _1 \gamma _2|
   \frac{ \alpha _1 \alpha _2^*+\alpha _2 \alpha_1^*}{ | \alpha _1 \alpha _2 |}
  J_1(4  | \alpha _1 \gamma _1 |)
   J_1(4 | \alpha _2 \gamma _2|)
  \Big )
\nonumber \\ &\times
   {\cal B}(\alpha_1,\alpha_2)\,  {\cal B}(\alpha_1+\gamma_1,\alpha_2+\gamma_2).
\label{eq:heq6} \end{align}
As in the single-particle case, we can divide the integral into
parts where the symbol $ {\cal B}$ is either replaced by 1 or by a step
function. We first consider the integral $I_{1,1}$, which is
given by Eq.~(\ref{eq:heq6}) with both factors of $ {\cal B}$
replaced by 1. We then can use a polar decomposition of the
complex variables $\alpha_i, \gamma_i$, whereby the angular
integrations are easily performed. For the integration over the modulus
of these variables we note that the integrand is only exponentially
suppressed in $|\gamma_i|$.  It is therefore prudent to first evaluate
the integration over $|\gamma_i|$, which results in
\begin{align}
  I_{1,1} &= 16 \int d |\alpha_1|  \, d |\alpha_2|\,
  e^{-2 ( | \alpha _1 | ^2+ | \alpha_2 | ^2)}  | \alpha _1 |
   | \alpha _2 |
\nonumber \\ & \hspace{5mm} \times
  (2  |\alpha _1|^2 +2  |   \alpha _2 | ^2-1).
\label{eq:heq7}  \end{align}
It is important to note here that the integration over $|\gamma_i|$
has resulted in exponential factors for the variables $|\alpha_i|$.
We will use this fact in the numerical evaluations presented below.
Eq.~(\ref{eq:heq7}) is easily evaluated and yields $I_{1,1}=1$. This
is because an operator with symbol 1 is equal to
the identity operator, which according to Eq.~(\ref{eq:unitDecomp})
will always yield  $\langle \hat{{\cal B}}^2
\rangle_\text{HV} =1$.

We now turn to the integral $I_{\theta,1}$,  which is
given by Eq.~(\ref{eq:heq6}) with $ {\cal B}(\alpha_1,\alpha_2)$ replaced by
$\theta(1-2|\alpha_1-\alpha_2|^2)$ and
$ {\cal B}(\alpha_1+\gamma_1,\alpha_2+\gamma_2)$
replaced by 1. The only difference to $I_{1,1}$ is the appearance of
the step function, which does not depend on $\gamma_1,\gamma_2$.
We can therefore perform the integration over $\gamma_i$ to obtain
\begin{align}
  I_{\theta,1} &= \frac{4}{\pi^2} \int d^2\alpha_1\, d^2\alpha_2\,
   e^{-2 |\alpha _1|^2-2 | \alpha _2 |^2}
  \Big ( 2 | \alpha _1| ^2+2 |
   \alpha _2| ^2
\nonumber \\ &\hspace{5mm}
  -2 \alpha _1 \alpha_2^*-2 \alpha _2 \alpha _1^*-1
  \Big)
   \theta \left(1-2 | \alpha _1-\alpha _2| ^2\right ).
\end{align}
Switching to relative coordinates $\sigma = \alpha_1+\alpha_2$ and
$\delta\alpha = \alpha_1-\alpha_2$ this integral can be evaluated and
yields
\begin{align}
    I_{\theta,1} &= 1 - \frac{ 2}{\sqrt{e}} \quad \approx -0.213 .
\end{align}

The remaining integral to evaluate
$\langle \hat{{\cal B}}^2  \rangle_\text{HV} $
is $I_{{\cal B},\theta}$,  which is
given by Eq.~(\ref{eq:heq6}) with $ {\cal B}(\alpha_1,\alpha_2)$ kept and
$ {\cal B}(\alpha_1+\gamma_1,\alpha_2+\gamma_2)$
replaced by the step function
$\theta(1-2|\alpha_1-\alpha_2+\gamma_1-\gamma_2|^2)$.
We were unable to solve this integral analytically. Even a numerical
evaluation is challenging because one has to perform an
eight-dimensional integration over a function that does not
rapidly converge to zero in the two variables $|\alpha_1|, |\alpha_2|$.
However, with the lessons learned in evaluating $I_{1,1}$ and
$I_{\theta , 1}$ we were able to numerically evaluate $I_{{\cal
    B},\theta}$ by using the following procedure.

\begin{figure}
\begin{center}
\includegraphics[width=6.5cm]{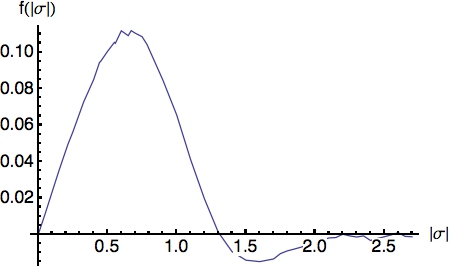}
\caption{\label{fig:numInt}
Intermediate numerical result for the evaluation of $I_{{\cal B},\theta}$.}
\end{center}
\end{figure}

(i) Using a suitable parametrization for the phases of the complex
variables $\alpha_i, \gamma_i$, it is possible to express the
integrand in a way that only depends on the relative phase of
$\alpha_1, \alpha_2$ (and not on their common phase). This
enables us to reduce the numerical integral to seven dimensions.
\\
(ii) Because large values of $|\gamma_i|$ are exponentially
suppressed, the step function effectively limits the relative
coordinate $\delta \alpha$ to small values. Hence, when
we use relative coordinates, the only variable that extends to
infinity
and in which the integrand
is not exponentially fast decreasing to zero is $|\sigma|$.
\\
(iii) In the evaluation of $I_{1,1}$ and $I_{\theta,1}$ we have seen
that integration over $\gamma_i$ results in an exponential factor
for $\alpha_i$. We may therefore expect a similar effect
may happen for $I_{{\cal B},\theta}$. The strategy to evaluate this
integral is therefore to first perform a six-dimensional
numerical integration over all phases as well as $|\gamma_i|$ and
$|\delta\alpha|$ to obtain a numerical function $f(|\sigma|)$,
which is displayed in Fig.~\ref{fig:numInt}.

To perform the six-dimensional numerical integration we have used
the NIntegrate function of
Mathematica\texttrademark with the adaptive Monte Carlo
method. To obtain Fig.~\ref{fig:numInt} we increased $|\sigma|$ in
steps of 0.05 from 0 to 2.7, with smaller step sizes around the
maximum of $f(|\sigma|)$.
For values of $|\sigma|$ that are larger than 3 it becomes impossible
to obtain accurate results, but within the limitations of the
numerical methods the results are consistent with a decreasing
function $f(|\sigma|)$.

The last step in the numerical evaluation is to numerically integrate
the function $f(|\sigma|)$. This can be done with standard methods
and yields $I_{{\cal B},\theta} \approx 0.078$. The full result for
the HV upper bound is then given by
\begin{align}
    \langle \hat{{\cal B}}^2  \rangle_\text{HV}
    &= I_{1,1}-2I_{\theta , 1} -2 I_{{\cal B},\theta}
  \; \; \approx 1.27 .
\end{align}

\end{appendix}

\bibliographystyle{apsrev4-1}
\bibliography{MoyalBell}

\begin{thebibliography}{45}%
\makeatletter
\providecommand \@ifxundefined [1]{%
 \@ifx{#1\undefined}
}%
\providecommand \@ifnum [1]{%
 \ifnum #1\expandafter \@firstoftwo
 \else \expandafter \@secondoftwo
 \fi
}%
\providecommand \@ifx [1]{%
 \ifx #1\expandafter \@firstoftwo
 \else \expandafter \@secondoftwo
 \fi
}%
\providecommand \natexlab [1]{#1}%
\providecommand \enquote  [1]{``#1''}%
\providecommand \bibnamefont  [1]{#1}%
\providecommand \bibfnamefont [1]{#1}%
\providecommand \citenamefont [1]{#1}%
\providecommand \href@noop [0]{\@secondoftwo}%
\providecommand \href [0]{\begingroup \@sanitize@url \@href}%
\providecommand \@href[1]{\@@startlink{#1}\@@href}%
\providecommand \@@href[1]{\endgroup#1\@@endlink}%
\providecommand \@sanitize@url [0]{\catcode `\\12\catcode `\$12\catcode
  `\&12\catcode `\#12\catcode `\^12\catcode `\_12\catcode `\%12\relax}%
\providecommand \@@startlink[1]{}%
\providecommand \@@endlink[0]{}%
\providecommand \url  [0]{\begingroup\@sanitize@url \@url }%
\providecommand \@url [1]{\endgroup\@href {#1}{\urlprefix }}%
\providecommand \urlprefix  [0]{URL }%
\providecommand \Eprint [0]{\href }%
\providecommand \doibase [0]{http://dx.doi.org/}%
\providecommand \selectlanguage [0]{\@gobble}%
\providecommand \bibinfo  [0]{\@secondoftwo}%
\providecommand \bibfield  [0]{\@secondoftwo}%
\providecommand \translation [1]{[#1]}%
\providecommand \BibitemOpen [0]{}%
\providecommand \bibitemStop [0]{}%
\providecommand \bibitemNoStop [0]{.\EOS\space}%
\providecommand \EOS [0]{\spacefactor3000\relax}%
\providecommand \BibitemShut  [1]{\csname bibitem#1\endcsname}%
\let\auto@bib@innerbib\@empty
\bibitem [{\citenamefont {Bell}(1964)}]{Bell:Physics1964}%
  \BibitemOpen
  \bibfield  {author} {\bibinfo {author} {\bibfnamefont {J.~S.}\ \bibnamefont
  {Bell}},\ }\href@noop {} {\bibfield  {journal} {\bibinfo  {journal}
  {Physics}\ }\textbf {\bibinfo {volume} {1}},\ \bibinfo {pages} {195}
  (\bibinfo {year} {1964})}\BibitemShut {NoStop}%
\bibitem [{\citenamefont {Clauser}\ \emph {et~al.}(1969)\citenamefont
  {Clauser}, \citenamefont {Horne}, \citenamefont {Shimony},\ and\
  \citenamefont {Holt}}]{PhysRevLett.23.880}%
  \BibitemOpen
  \bibfield  {author} {\bibinfo {author} {\bibfnamefont {J.~F.}\ \bibnamefont
  {Clauser}}, \bibinfo {author} {\bibfnamefont {M.~A.}\ \bibnamefont {Horne}},
  \bibinfo {author} {\bibfnamefont {A.}~\bibnamefont {Shimony}}, \ and\
  \bibinfo {author} {\bibfnamefont {R.~A.}\ \bibnamefont {Holt}},\ }\href
  {\doibase 10.1103/PhysRevLett.23.880} {\bibfield  {journal} {\bibinfo
  {journal} {Phys. Rev. Lett.}\ }\textbf {\bibinfo {volume} {23}},\ \bibinfo
  {pages} {880} (\bibinfo {year} {1969})}\BibitemShut {NoStop}%
\bibitem [{\citenamefont {Freedman}\ and\ \citenamefont
  {Clauser}(1972)}]{PhysRevLett.28.938}%
  \BibitemOpen
  \bibfield  {author} {\bibinfo {author} {\bibfnamefont {S.~J.}\ \bibnamefont
  {Freedman}}\ and\ \bibinfo {author} {\bibfnamefont {J.~F.}\ \bibnamefont
  {Clauser}},\ }\href {\doibase 10.1103/PhysRevLett.28.938} {\bibfield
  {journal} {\bibinfo  {journal} {Phys. Rev. Lett.}\ }\textbf {\bibinfo
  {volume} {28}},\ \bibinfo {pages} {938} (\bibinfo {year} {1972})}\BibitemShut
  {NoStop}%
\bibitem [{\citenamefont {Aspect}\ \emph {et~al.}(1982)\citenamefont {Aspect},
  \citenamefont {Dalibard},\ and\ \citenamefont {Roger}}]{PhysRevLett.49.1804}%
  \BibitemOpen
  \bibfield  {author} {\bibinfo {author} {\bibfnamefont {A.}~\bibnamefont
  {Aspect}}, \bibinfo {author} {\bibfnamefont {J.}~\bibnamefont {Dalibard}}, \
  and\ \bibinfo {author} {\bibfnamefont {G.}~\bibnamefont {Roger}},\ }\href
  {\doibase 10.1103/PhysRevLett.49.1804} {\bibfield  {journal} {\bibinfo
  {journal} {Phys. Rev. Lett.}\ }\textbf {\bibinfo {volume} {49}},\ \bibinfo
  {pages} {1804} (\bibinfo {year} {1982})}\BibitemShut {NoStop}%
\bibitem [{\citenamefont {Weihs}\ \emph {et~al.}(1998)\citenamefont {Weihs},
  \citenamefont {Jennewein}, \citenamefont {Simon}, \citenamefont
  {Weinfurter},\ and\ \citenamefont {Zeilinger}}]{PhysRevLett.81.5039}%
  \BibitemOpen
  \bibfield  {author} {\bibinfo {author} {\bibfnamefont {G.}~\bibnamefont
  {Weihs}}, \bibinfo {author} {\bibfnamefont {T.}~\bibnamefont {Jennewein}},
  \bibinfo {author} {\bibfnamefont {C.}~\bibnamefont {Simon}}, \bibinfo
  {author} {\bibfnamefont {H.}~\bibnamefont {Weinfurter}}, \ and\ \bibinfo
  {author} {\bibfnamefont {A.}~\bibnamefont {Zeilinger}},\ }\href {\doibase
  10.1103/PhysRevLett.81.5039} {\bibfield  {journal} {\bibinfo  {journal}
  {Phys. Rev. Lett.}\ }\textbf {\bibinfo {volume} {81}},\ \bibinfo {pages}
  {5039} (\bibinfo {year} {1998})}\BibitemShut {NoStop}%
\bibitem [{\citenamefont {Giustina}\ \emph {et~al.}(2013)\citenamefont
  {Giustina}, \citenamefont {Mech}, \citenamefont {Ramelow}, \citenamefont
  {Wittmann}, \citenamefont {Kofler}, \citenamefont {Beyer}, \citenamefont
  {Lita}, \citenamefont {Calkins}, \citenamefont {Gerrits}, \citenamefont
  {Nam}, \citenamefont {Ursin},\ and\ \citenamefont
  {Zeilinger}}]{BellLoopholeZeilinger2013}%
  \BibitemOpen
  \bibfield  {author} {\bibinfo {author} {\bibfnamefont {M.}~\bibnamefont
  {Giustina}}, \bibinfo {author} {\bibfnamefont {A.}~\bibnamefont {Mech}},
  \bibinfo {author} {\bibfnamefont {S.}~\bibnamefont {Ramelow}}, \bibinfo
  {author} {\bibfnamefont {B.}~\bibnamefont {Wittmann}}, \bibinfo {author}
  {\bibfnamefont {J.}~\bibnamefont {Kofler}}, \bibinfo {author} {\bibfnamefont
  {J.}~\bibnamefont {Beyer}}, \bibinfo {author} {\bibfnamefont
  {A.}~\bibnamefont {Lita}}, \bibinfo {author} {\bibfnamefont {B.}~\bibnamefont
  {Calkins}}, \bibinfo {author} {\bibfnamefont {T.}~\bibnamefont {Gerrits}},
  \bibinfo {author} {\bibfnamefont {S.~W.}\ \bibnamefont {Nam}}, \bibinfo
  {author} {\bibfnamefont {R.}~\bibnamefont {Ursin}}, \ and\ \bibinfo {author}
  {\bibfnamefont {A.}~\bibnamefont {Zeilinger}},\ }\href {\doibase
  10.1038/nature12012} {\bibfield  {journal} {\bibinfo  {journal} {Nature}\
  }\textbf {\bibinfo {volume} {2013/04/14/online}},\ \bibinfo {pages} {4}
  (\bibinfo {year} {2013})}\BibitemShut {NoStop}%
\bibitem [{\citenamefont {Einstein}\ \emph {et~al.}(1935)\citenamefont
  {Einstein}, \citenamefont {Podolsky},\ and\ \citenamefont
  {Rosen}}]{PhysRev.47.777}%
  \BibitemOpen
  \bibfield  {author} {\bibinfo {author} {\bibfnamefont {A.}~\bibnamefont
  {Einstein}}, \bibinfo {author} {\bibfnamefont {B.}~\bibnamefont {Podolsky}},
  \ and\ \bibinfo {author} {\bibfnamefont {N.}~\bibnamefont {Rosen}},\ }\href
  {\doibase 10.1103/PhysRev.47.777} {\bibfield  {journal} {\bibinfo  {journal}
  {Phys. Rev.}\ }\textbf {\bibinfo {volume} {47}},\ \bibinfo {pages} {777}
  (\bibinfo {year} {1935})}\BibitemShut {NoStop}%
\bibitem [{\citenamefont {Reid}\ \emph {et~al.}(2009)\citenamefont {Reid},
  \citenamefont {Drummond}, \citenamefont {Bowen}, \citenamefont {Cavalcanti},
  \citenamefont {Lam}, \citenamefont {Bachor}, \citenamefont {Andersen},\ and\
  \citenamefont {Leuchs}}]{RevModPhys.81.1727}%
  \BibitemOpen
  \bibfield  {author} {\bibinfo {author} {\bibfnamefont {M.~D.}\ \bibnamefont
  {Reid}}, \bibinfo {author} {\bibfnamefont {P.~D.}\ \bibnamefont {Drummond}},
  \bibinfo {author} {\bibfnamefont {W.~P.}\ \bibnamefont {Bowen}}, \bibinfo
  {author} {\bibfnamefont {E.~G.}\ \bibnamefont {Cavalcanti}}, \bibinfo
  {author} {\bibfnamefont {P.~K.}\ \bibnamefont {Lam}}, \bibinfo {author}
  {\bibfnamefont {H.~A.}\ \bibnamefont {Bachor}}, \bibinfo {author}
  {\bibfnamefont {U.~L.}\ \bibnamefont {Andersen}}, \ and\ \bibinfo {author}
  {\bibfnamefont {G.}~\bibnamefont {Leuchs}},\ }\href@noop {} {\bibfield
  {journal} {\bibinfo  {journal} {Rev. Mod. Phys.}\ }\textbf {\bibinfo {volume}
  {81}},\ \bibinfo {pages} {1727} (\bibinfo {year} {2009})}\BibitemShut
  {NoStop}%
\bibitem [{\citenamefont {Leonhardt}\ and\ \citenamefont
  {Vaccaro}(1995)}]{JModOpt42-939}%
  \BibitemOpen
  \bibfield  {author} {\bibinfo {author} {\bibfnamefont {U.}~\bibnamefont
  {Leonhardt}}\ and\ \bibinfo {author} {\bibfnamefont {J.~A.}\ \bibnamefont
  {Vaccaro}},\ }\href {\doibase 10.1080/09500349514550851} {\bibfield
  {journal} {\bibinfo  {journal} {J. Mod. Opt.}\ }\textbf {\bibinfo {volume}
  {42}},\ \bibinfo {pages} {939 } (\bibinfo {year} {1995})}\BibitemShut
  {NoStop}%
\bibitem [{\citenamefont {Gour}\ \emph {et~al.}(2004)\citenamefont {Gour},
  \citenamefont {Khanna}, \citenamefont {Mann},\ and\ \citenamefont
  {Revzen}}]{Gour2004415}%
  \BibitemOpen
  \bibfield  {author} {\bibinfo {author} {\bibfnamefont {G.}~\bibnamefont
  {Gour}}, \bibinfo {author} {\bibfnamefont {F.~C.}\ \bibnamefont {Khanna}},
  \bibinfo {author} {\bibfnamefont {A.}~\bibnamefont {Mann}}, \ and\ \bibinfo
  {author} {\bibfnamefont {M.}~\bibnamefont {Revzen}},\ }\href {\doibase DOI:
  10.1016/j.physleta.2004.03.018} {\bibfield  {journal} {\bibinfo  {journal}
  {Phys. Lett. A}\ }\textbf {\bibinfo {volume} {324}},\ \bibinfo {pages} {415 }
  (\bibinfo {year} {2004})}\BibitemShut {NoStop}%
\bibitem [{\citenamefont {Praxmeyer}\ \emph {et~al.}(2005)\citenamefont
  {Praxmeyer}, \citenamefont {Englert},\ and\ \citenamefont
  {W\'{o}dkiewicz}}]{Praxmeyer:EPJD2005}%
  \BibitemOpen
  \bibfield  {author} {\bibinfo {author} {\bibfnamefont {L.}~\bibnamefont
  {Praxmeyer}}, \bibinfo {author} {\bibfnamefont {B.-G.}\ \bibnamefont
  {Englert}}, \ and\ \bibinfo {author} {\bibfnamefont {K.}~\bibnamefont
  {W\'{o}dkiewicz}},\ }\href {\doibase 10.1140/epjd/e2005-00021-1} {\bibfield
  {journal} {\bibinfo  {journal} {Europ. Phys. J. D}\ }\textbf {\bibinfo
  {volume} {32}},\ \bibinfo {pages} {227} (\bibinfo {year} {2005})}\BibitemShut
  {NoStop}%
\bibitem [{\citenamefont {Cavalcanti}\ \emph {et~al.}(2007)\citenamefont
  {Cavalcanti}, \citenamefont {Foster}, \citenamefont {Reid},\ and\
  \citenamefont {Drummond}}]{PhysRevLett.99.210405}%
  \BibitemOpen
  \bibfield  {author} {\bibinfo {author} {\bibfnamefont {E.~G.}\ \bibnamefont
  {Cavalcanti}}, \bibinfo {author} {\bibfnamefont {C.~J.}\ \bibnamefont
  {Foster}}, \bibinfo {author} {\bibfnamefont {M.~D.}\ \bibnamefont {Reid}}, \
  and\ \bibinfo {author} {\bibfnamefont {P.~D.}\ \bibnamefont {Drummond}},\
  }\href {\doibase 10.1103/PhysRevLett.99.210405} {\bibfield  {journal}
  {\bibinfo  {journal} {Phys. Rev. Lett.}\ }\textbf {\bibinfo {volume} {99}},\
  \bibinfo {pages} {210405} (\bibinfo {year} {2007})}\BibitemShut {NoStop}%
\bibitem [{\citenamefont {He}\ \emph {et~al.}(2010)\citenamefont {He},
  \citenamefont {Cavalcanti}, \citenamefont {Reid},\ and\ \citenamefont
  {Drummond}}]{arXiv:1005.2208}%
  \BibitemOpen
  \bibfield  {author} {\bibinfo {author} {\bibfnamefont {Q.~Y.}\ \bibnamefont
  {He}}, \bibinfo {author} {\bibfnamefont {E.~G.}\ \bibnamefont {Cavalcanti}},
  \bibinfo {author} {\bibfnamefont {M.~D.}\ \bibnamefont {Reid}}, \ and\
  \bibinfo {author} {\bibfnamefont {P.~D.}\ \bibnamefont {Drummond}},\ }\href
  {\doibase 10.1103/PhysRevA.81.062106} {\bibfield  {journal} {\bibinfo
  {journal} {Phys. Rev. A}\ }\textbf {\bibinfo {volume} {81}},\ \bibinfo
  {pages} {062106} (\bibinfo {year} {2010})}\BibitemShut {NoStop}%
\bibitem [{\citenamefont {Malley}(2004)}]{PhysRevA.69.022118}%
  \BibitemOpen
  \bibfield  {author} {\bibinfo {author} {\bibfnamefont {J.~D.}\ \bibnamefont
  {Malley}},\ }\href {\doibase 10.1103/PhysRevA.69.022118} {\bibfield
  {journal} {\bibinfo  {journal} {Phys. Rev. A}\ }\textbf {\bibinfo {volume}
  {69}},\ \bibinfo {pages} {022118} (\bibinfo {year} {2004})}\BibitemShut
  {NoStop}%
\bibitem [{\citenamefont {Malley}\ and\ \citenamefont
  {Fine}(2005)}]{Malley200551}%
  \BibitemOpen
  \bibfield  {author} {\bibinfo {author} {\bibfnamefont {J.~D.}\ \bibnamefont
  {Malley}}\ and\ \bibinfo {author} {\bibfnamefont {A.}~\bibnamefont {Fine}},\
  }\href {\doibase DOI: 10.1016/j.physleta.2005.06.032} {\bibfield  {journal}
  {\bibinfo  {journal} {Phys. Lett. A}\ }\textbf {\bibinfo {volume} {347}},\
  \bibinfo {pages} {51 } (\bibinfo {year} {2005})}\BibitemShut {NoStop}%
\bibitem [{\citenamefont {Banaszek}\ and\ \citenamefont
  {W\'odkiewicz}(1999)}]{PhysRevLett.82.2009}%
  \BibitemOpen
  \bibfield  {author} {\bibinfo {author} {\bibfnamefont {K.}~\bibnamefont
  {Banaszek}}\ and\ \bibinfo {author} {\bibfnamefont {K.}~\bibnamefont
  {W\'odkiewicz}},\ }\href {\doibase 10.1103/PhysRevLett.82.2009} {\bibfield
  {journal} {\bibinfo  {journal} {Phys. Rev. Lett.}\ }\textbf {\bibinfo
  {volume} {82}},\ \bibinfo {pages} {2009} (\bibinfo {year}
  {1999})}\BibitemShut {NoStop}%
\bibitem [{\citenamefont {Banaszek}\ and\ \citenamefont
  {W\'odkiewicz}(1998)}]{PhysRevA.58.4345}%
  \BibitemOpen
  \bibfield  {author} {\bibinfo {author} {\bibfnamefont {K.}~\bibnamefont
  {Banaszek}}\ and\ \bibinfo {author} {\bibfnamefont {K.}~\bibnamefont
  {W\'odkiewicz}},\ }\href {\doibase 10.1103/PhysRevA.58.4345} {\bibfield
  {journal} {\bibinfo  {journal} {Phys. Rev. A}\ }\textbf {\bibinfo {volume}
  {58}},\ \bibinfo {pages} {4345} (\bibinfo {year} {1998})}\BibitemShut
  {NoStop}%
\bibitem [{\citenamefont {Revzen}\ \emph {et~al.}(2005)\citenamefont {Revzen},
  \citenamefont {Mello}, \citenamefont {Mann},\ and\ \citenamefont
  {Johansen}}]{revzen:022103}%
  \BibitemOpen
  \bibfield  {author} {\bibinfo {author} {\bibfnamefont {M.}~\bibnamefont
  {Revzen}}, \bibinfo {author} {\bibfnamefont {P.~A.}\ \bibnamefont {Mello}},
  \bibinfo {author} {\bibfnamefont {A.}~\bibnamefont {Mann}}, \ and\ \bibinfo
  {author} {\bibfnamefont {L.~M.}\ \bibnamefont {Johansen}},\ }\href {\doibase
  10.1103/PhysRevA.71.022103} {\bibfield  {journal} {\bibinfo  {journal} {Phys.
  Rev. A}\ }\textbf {\bibinfo {volume} {71}},\ \bibinfo {eid} {022103}
  (\bibinfo {year} {2005})}\BibitemShut {NoStop}%
\bibitem [{\citenamefont {Ozorio~de Almeida}(2009)}]{de2009entanglement}%
  \BibitemOpen
  \bibfield  {author} {\bibinfo {author} {\bibfnamefont {A.}~\bibnamefont
  {Ozorio~de Almeida}},\ }in\ \href@noop {} {\emph {\bibinfo {booktitle}
  {Entanglement and Decoherence. Foundations and modern trends}}},\ Vol.\
  \bibinfo {volume} {768}\ (\bibinfo  {publisher} {Springer},\ \bibinfo
  {address} {Berlin},\ \bibinfo {year} {2009})\ pp.\ \bibinfo {pages}
  {157--219}\BibitemShut {NoStop}%
\bibitem [{\citenamefont {Wigner}(1932)}]{PhysRev.40.749}%
  \BibitemOpen
  \bibfield  {author} {\bibinfo {author} {\bibfnamefont {E.}~\bibnamefont
  {Wigner}},\ }\href {\doibase 10.1103/PhysRev.40.749} {\bibfield  {journal}
  {\bibinfo  {journal} {Phys. Rev.}\ }\textbf {\bibinfo {volume} {40}},\
  \bibinfo {pages} {749} (\bibinfo {year} {1932})}\BibitemShut {NoStop}%
\bibitem [{\citenamefont {Folland}(1989)}]{Foll89}%
  \BibitemOpen
  \bibfield  {author} {\bibinfo {author} {\bibfnamefont {G.~B.}\ \bibnamefont
  {Folland}},\ }\href@noop {} {\emph {\bibinfo {title} {Harmonic Analysis in
  Phase Space}}}\ (\bibinfo  {publisher} {Princeton University Press},\
  \bibinfo {address} {Princeton},\ \bibinfo {year} {1989})\BibitemShut
  {NoStop}%
\bibitem [{\citenamefont {Royer}(1977)}]{Roy77}%
  \BibitemOpen
  \bibfield  {author} {\bibinfo {author} {\bibfnamefont {A.}~\bibnamefont
  {Royer}},\ }\href@noop {} {\bibfield  {journal} {\bibinfo  {journal} {Phys.
  Rev. A}\ }\textbf {\bibinfo {volume} {15}},\ \bibinfo {pages} {449} (\bibinfo
  {year} {1977})}\BibitemShut {NoStop}%
\bibitem [{\citenamefont {Grossmann}(1976)}]{Gro76}%
  \BibitemOpen
  \bibfield  {author} {\bibinfo {author} {\bibfnamefont {A.}~\bibnamefont
  {Grossmann}},\ }\href@noop {} {\bibfield  {journal} {\bibinfo  {journal}
  {Commun. Math. Phys.}\ }\textbf {\bibinfo {volume} {48}},\ \bibinfo {pages}
  {191} (\bibinfo {year} {1976})}\BibitemShut {NoStop}%
\bibitem [{\citenamefont {{M.~V.~Karasev and T.~A.~Osborn}}(2004)}]{KO3}%
  \BibitemOpen
  \bibfield  {author} {\bibinfo {author} {\bibnamefont {{M.~V.~Karasev and
  T.~A.~Osborn}}},\ }\href {\doibase 10.1103/PhysRevLett.83.3758} {\bibfield
  {journal} {\bibinfo  {journal} {J. Phys. A.}\ }\textbf {\bibinfo {volume}
  {37}},\ \bibinfo {pages} {2345} (\bibinfo {year} {2004})}\BibitemShut
  {NoStop}%
\bibitem [{\citenamefont {Revzen}(2006)}]{FoundPhys36-546}%
  \BibitemOpen
  \bibfield  {author} {\bibinfo {author} {\bibfnamefont {M.}~\bibnamefont
  {Revzen}},\ }\href {\doibase 10.1007/s10701-005-9037-5} {\bibfield  {journal}
  {\bibinfo  {journal} {Found. Phys.}\ }\textbf {\bibinfo {volume} {36}},\
  \bibinfo {pages} {546} (\bibinfo {year} {2006})}\BibitemShut {NoStop}%
\bibitem [{\citenamefont {Kalev}\ \emph {et~al.}(2009)\citenamefont {Kalev},
  \citenamefont {Mann}, \citenamefont {Mello},\ and\ \citenamefont
  {Revzen}}]{PhysRevA.79.014104}%
  \BibitemOpen
  \bibfield  {author} {\bibinfo {author} {\bibfnamefont {A.}~\bibnamefont
  {Kalev}}, \bibinfo {author} {\bibfnamefont {A.}~\bibnamefont {Mann}},
  \bibinfo {author} {\bibfnamefont {P.~A.}\ \bibnamefont {Mello}}, \ and\
  \bibinfo {author} {\bibfnamefont {M.}~\bibnamefont {Revzen}},\ }\href
  {\doibase 10.1103/PhysRevA.79.014104} {\bibfield  {journal} {\bibinfo
  {journal} {Phys. Rev. A}\ }\textbf {\bibinfo {volume} {79}},\ \bibinfo
  {pages} {014104} (\bibinfo {year} {2009})}\BibitemShut {NoStop}%
\bibitem [{\citenamefont {Kenfack}\ and\ \citenamefont
  {\.{Z}yczkowski}(2004)}]{1464-4266-6-10-003}%
  \BibitemOpen
  \bibfield  {author} {\bibinfo {author} {\bibfnamefont {A.}~\bibnamefont
  {Kenfack}}\ and\ \bibinfo {author} {\bibfnamefont {K.}~\bibnamefont
  {\.{Z}yczkowski}},\ }\href {http://stacks.iop.org/1464-4266/6/i=10/a=003}
  {\bibfield  {journal} {\bibinfo  {journal} {J. Opt. B: Quantum Semicl.}\
  }\textbf {\bibinfo {volume} {6}},\ \bibinfo {pages} {396} (\bibinfo {year}
  {2004})}\BibitemShut {NoStop}%
\bibitem [{\citenamefont {Benedict}\ and\ \citenamefont
  {Czirj\'ak}(1999)}]{PhysRevA.60.4034}%
  \BibitemOpen
  \bibfield  {author} {\bibinfo {author} {\bibfnamefont {M.~G.}\ \bibnamefont
  {Benedict}}\ and\ \bibinfo {author} {\bibfnamefont {A.}~\bibnamefont
  {Czirj\'ak}},\ }\href {\doibase 10.1103/PhysRevA.60.4034} {\bibfield
  {journal} {\bibinfo  {journal} {Phys. Rev. A}\ }\textbf {\bibinfo {volume}
  {60}},\ \bibinfo {pages} {4034} (\bibinfo {year} {1999})}\BibitemShut
  {NoStop}%
\bibitem [{\citenamefont {Redhead}(1987)}]{Redhead:Incompleteness}%
  \BibitemOpen
  \bibfield  {author} {\bibinfo {author} {\bibfnamefont {M.}~\bibnamefont
  {Redhead}},\ }\href@noop {} {\emph {\bibinfo {title} {Incompleteness,
  Nonlocality and Realism}}}\ (\bibinfo  {publisher} {Clarendon},\ \bibinfo
  {year} {1987})\BibitemShut {NoStop}%
\bibitem [{\citenamefont {Malley}(1998)}]{PhysRevA.58.812}%
  \BibitemOpen
  \bibfield  {author} {\bibinfo {author} {\bibfnamefont {J.~D.}\ \bibnamefont
  {Malley}},\ }\href {\doibase 10.1103/PhysRevA.58.812} {\bibfield  {journal}
  {\bibinfo  {journal} {Phys. Rev. A}\ }\textbf {\bibinfo {volume} {58}},\
  \bibinfo {pages} {812} (\bibinfo {year} {1998})}\BibitemShut {NoStop}%
\bibitem [{\citenamefont {Fine}(1982{\natexlab{a}})}]{fine1982hidden}%
  \BibitemOpen
  \bibfield  {author} {\bibinfo {author} {\bibfnamefont {A.}~\bibnamefont
  {Fine}},\ }\href@noop {} {\bibfield  {journal} {\bibinfo  {journal} {Phys.
  Rev. Lett.}\ }\textbf {\bibinfo {volume} {48}},\ \bibinfo {pages} {291}
  (\bibinfo {year} {1982}{\natexlab{a}})}\BibitemShut {NoStop}%
\bibitem [{\citenamefont {Bobo}(2010)}]{Bobo2010}%
  \BibitemOpen
  \bibfield  {author} {\bibinfo {author} {\bibfnamefont {I.~G.}\ \bibnamefont
  {Bobo}},\ }\href {http://eprints.ucm.es/10636/1/T31869.pdf} {\emph {\bibinfo
  {title} {Quantum conditional probability: implications for conceptual change
  of science}}}\ (\bibinfo  {publisher} {Universidad Complutense de Madrid},\
  \bibinfo {year} {2010})\ \bibinfo {note} {iSBN:
  978-84-693-3483-6}\BibitemShut {NoStop}%
\bibitem [{\citenamefont {Beltrametti}\ and\ \citenamefont
  {Cassinelli}(1981)}]{Beltrametti:QM81}%
  \BibitemOpen
  \bibfield  {author} {\bibinfo {author} {\bibfnamefont {E.}~\bibnamefont
  {Beltrametti}}\ and\ \bibinfo {author} {\bibfnamefont {G.}~\bibnamefont
  {Cassinelli}},\ }\href@noop {} {\emph {\bibinfo {title} {The Logic of Quantum
  Mechanics}}}\ (\bibinfo  {publisher} {Addison -Wesley},\ \bibinfo {year}
  {1981})\BibitemShut {NoStop}%
\bibitem [{\citenamefont {Gleason}(1957)}]{gleason1957measures}%
  \BibitemOpen
  \bibfield  {author} {\bibinfo {author} {\bibfnamefont {A.~M.}\ \bibnamefont
  {Gleason}},\ }\href@noop {} {\bibfield  {journal} {\bibinfo  {journal} {J.
  Math. Mech}\ }\textbf {\bibinfo {volume} {6}},\ \bibinfo {pages} {885}
  (\bibinfo {year} {1957})}\BibitemShut {NoStop}%
\bibitem [{\citenamefont {Busch}\ and\ \citenamefont
  {Lahti}(1996)}]{jmp/37/6/10.1063/1.531530}%
  \BibitemOpen
  \bibfield  {author} {\bibinfo {author} {\bibfnamefont {P.}~\bibnamefont
  {Busch}}\ and\ \bibinfo {author} {\bibfnamefont {P.~J.}\ \bibnamefont
  {Lahti}},\ }\href {\doibase http://dx.doi.org/10.1063/1.531530} {\bibfield
  {journal} {\bibinfo  {journal} {J. Math. Phys.}\ }\textbf {\bibinfo {volume}
  {37}},\ \bibinfo {pages} {2585} (\bibinfo {year} {1996})}\BibitemShut
  {NoStop}%
\bibitem [{\citenamefont {Hemmick}\ and\ \citenamefont
  {Shakur}(2012)}]{Hemmick2012}%
  \BibitemOpen
  \bibfield  {author} {\bibinfo {author} {\bibfnamefont {D.~L.}\ \bibnamefont
  {Hemmick}}\ and\ \bibinfo {author} {\bibfnamefont {A.~M.}\ \bibnamefont
  {Shakur}},\ }\href@noop {} {\emph {\bibinfo {title} {Bell's Theorem and
  Quantum Realism}}}\ (\bibinfo  {publisher} {Springer},\ \bibinfo {address}
  {Heidelberg},\ \bibinfo {year} {2012})\BibitemShut {NoStop}%
\bibitem [{\citenamefont {Gudder}(1979)}]{gudder1979stochastic}%
  \BibitemOpen
  \bibfield  {author} {\bibinfo {author} {\bibfnamefont {S.}~\bibnamefont
  {Gudder}},\ }\href@noop {} {\emph {\bibinfo {title} {Stochastic methods in
  quantum mechanics}}}\ (\bibinfo  {publisher} {North Holland},\ \bibinfo
  {address} {New York},\ \bibinfo {year} {1979})\BibitemShut {NoStop}%
\bibitem [{\citenamefont {Fine}(1982{\natexlab{b}})}]{fine1982joint}%
  \BibitemOpen
  \bibfield  {author} {\bibinfo {author} {\bibfnamefont {A.}~\bibnamefont
  {Fine}},\ }\href@noop {} {\bibfield  {journal} {\bibinfo  {journal} {J. Math.
  Phys.}\ }\textbf {\bibinfo {volume} {23}},\ \bibinfo {pages} {1306} (\bibinfo
  {year} {1982}{\natexlab{b}})}\BibitemShut {NoStop}%
\bibitem [{\citenamefont {Bell}(1966)}]{bell1966problem}%
  \BibitemOpen
  \bibfield  {author} {\bibinfo {author} {\bibfnamefont {J.~S.}\ \bibnamefont
  {Bell}},\ }\href@noop {} {\bibfield  {journal} {\bibinfo  {journal} {Reviews
  of Modern Physics}\ }\textbf {\bibinfo {volume} {38}},\ \bibinfo {pages}
  {447} (\bibinfo {year} {1966})}\BibitemShut {NoStop}%
\bibitem [{\citenamefont {Kochen}\ and\ \citenamefont {Specker}(1968)}]{Koc}%
  \BibitemOpen
  \bibfield  {author} {\bibinfo {author} {\bibfnamefont {S.}~\bibnamefont
  {Kochen}}\ and\ \bibinfo {author} {\bibfnamefont {E.}~\bibnamefont
  {Specker}},\ }\href@noop {} {\bibfield  {journal} {\bibinfo  {journal}
  {Indiana Univ. Math. J.}\ }\textbf {\bibinfo {volume} {17}},\ \bibinfo
  {pages} {59} (\bibinfo {year} {1968})}\BibitemShut {NoStop}%
\bibitem [{\citenamefont {Bracken}\ \emph {et~al.}(1999)\citenamefont
  {Bracken}, \citenamefont {Doebner},\ and\ \citenamefont
  {Wood}}]{PhysRevLett.83.3758}%
  \BibitemOpen
  \bibfield  {author} {\bibinfo {author} {\bibfnamefont {A.~J.}\ \bibnamefont
  {Bracken}}, \bibinfo {author} {\bibfnamefont {H.-D.}\ \bibnamefont
  {Doebner}}, \ and\ \bibinfo {author} {\bibfnamefont {J.~G.}\ \bibnamefont
  {Wood}},\ }\href {\doibase 10.1103/PhysRevLett.83.3758} {\bibfield  {journal}
  {\bibinfo  {journal} {Phys. Rev. Lett.}\ }\textbf {\bibinfo {volume} {83}},\
  \bibinfo {pages} {3758} (\bibinfo {year} {1999})}\BibitemShut {NoStop}%
\bibitem [{\citenamefont {Abramowitz}\ and\ \citenamefont
  {Stegun}(1964)}]{abramowitz}%
  \BibitemOpen
  \bibfield  {author} {\bibinfo {author} {\bibfnamefont {M.}~\bibnamefont
  {Abramowitz}}\ and\ \bibinfo {author} {\bibfnamefont {I.~A.}\ \bibnamefont
  {Stegun}},\ }\href@noop {} {\emph {\bibinfo {title} {Handbook of mathematical
  functions}}}\ (\bibinfo  {publisher} {National Bureau of Standards},\
  \bibinfo {address} {Washington, D.C.},\ \bibinfo {year} {1964})\BibitemShut
  {NoStop}%
\bibitem [{\citenamefont {Cahill}\ and\ \citenamefont
  {Glauber}(1969)}]{PhysRev.177.1857}%
  \BibitemOpen
  \bibfield  {author} {\bibinfo {author} {\bibfnamefont {K.~E.}\ \bibnamefont
  {Cahill}}\ and\ \bibinfo {author} {\bibfnamefont {R.~J.}\ \bibnamefont
  {Glauber}},\ }\href {\doibase 10.1103/PhysRev.177.1857} {\bibfield  {journal}
  {\bibinfo  {journal} {Phys. Rev.}\ }\textbf {\bibinfo {volume} {177}},\
  \bibinfo {pages} {1857} (\bibinfo {year} {1969})}\BibitemShut {NoStop}%
\bibitem [{\citenamefont {Gradshteyn}\ and\ \citenamefont
  {Ryzhik}(2007)}]{GradshteynRyzhik}%
  \BibitemOpen
  \bibfield  {author} {\bibinfo {author} {\bibnamefont {Gradshteyn}}\ and\
  \bibinfo {author} {\bibnamefont {Ryzhik}},\ }\href@noop {} {\emph {\bibinfo
  {title} {Table of Integrals, Series, and Products, 7th Edition}}},\ edited
  by\ \bibinfo {editor} {\bibfnamefont {A.}~\bibnamefont {Jeffrey}}\ and\
  \bibinfo {editor} {\bibfnamefont {D.}~\bibnamefont {Zwillinger}}\ (\bibinfo
  {publisher} {Academic Press},\ \bibinfo {year} {2007})\BibitemShut {NoStop}%
\bibitem [{\citenamefont {Braunstein}\ and\ \citenamefont
  {Caves}(1990)}]{Braunstein199022}%
  \BibitemOpen
  \bibfield  {author} {\bibinfo {author} {\bibfnamefont {S.~L.}\ \bibnamefont
  {Braunstein}}\ and\ \bibinfo {author} {\bibfnamefont {C.~M.}\ \bibnamefont
  {Caves}},\ }\href {\doibase 10.1016/0003-4916(90)90339-P} {\bibfield
  {journal} {\bibinfo  {journal} {Ann. Phys.}\ }\textbf {\bibinfo {volume}
  {202}},\ \bibinfo {pages} {22 } (\bibinfo {year} {1990})}\BibitemShut
  {NoStop}%
\end{thebibliography}%
\end{document}